\documentclass[manuscript]{aastex}
\usepackage{lscape}

\slugcomment{}

\shorttitle{Elements in Low-redshift Sub-DLAs}
\shortauthors{Som et al.}

\begin{document}

\title{Hubble Space Telescope Observations of Sub-Damped Lyman-alpha Absorbers at $z < 0.5$, and Implications for 
Galaxy Chemical Evolution}

\author{Debopam Som and Varsha P. Kulkarni}
\affil{Univ. of South Carolina, Dept. of Physics \& Astronomy, Columbia, SC 29208}
\email{som@email.sc.edu, kulkarni@sc.edu}

\author{Joseph Meiring\altaffilmark{1}}
\affil{Univ. of Massachusetts, Dept. of Astronomy, Amherst, MA 01003 }

\author{Donald G. York\altaffilmark{2}}
\affil{Univ. of Chicago, Dept. of Astronomy \& Astrophysics, Chicago, IL 60637}

\author{Celine P\'eroux}
\affil{Aix Marseille Universit\'e, CNRS, Laboratoire d'Astrophysique de Marseille, UMR 7326, 13388, Marseille, France}

\author{James T. Lauroesch}
\affil{Univ. of Louisville, Dept. of Physics \& Astronomy, Louisville, KY 40292}

\author{Monique C. Aller}
\affil{Univ. of South Carolina, Dept. of Physics \& Astronomy, Columbia, SC 29208}

\author{Pushpa Khare}
\affil{Inter-University Center for Astronomy \& Astrophysics, Pune 411007, India}

\altaffiltext{1}{Current affiliation: Exigo Energy Solutions}
\altaffiltext{2}{Also, the Enrico Fermi Institute}

\begin{abstract}

We report observations of four sub-damped Lyman-alpha (sub-DLA) quasar absorbers at $z<0.5$ obtained with the Hubble Space Telescope Cosmic Origins Spectrograph. We measure the available
neutrals or ions of C, N, O, Si, P, S, Ar, Mn, Fe, and/or Ni. Our data have doubled the sub-DLA metallicity samples at $z<0.5$ and improved constraints on sub-DLA chemical evolution. All four
of our sub-DLAs are consistent with near-solar or super-solar metallicities and relatively modest ionization corrections; observations of more lines and detailed modeling will help to verify
this. Combining our data with measurements from the literature, we confirm previous suggestions that the N(HI)-weighted mean metallicity of sub-DLAs exceeds that of DLAs at all redshifts
studied, even after making ionization corrections for sub-DLAs. The absorber toward PHL 1598 shows significant dust depletion. The absorbers toward PHL 1226 and PKS 0439-433 show the S/P
ratio consistent with solar, i.e., they lack a profound odd-even effect. The absorber toward Q0439-433 shows super-solar Mn/Fe. For several sub-DLAs at $z<0.5$, [N/S] is below the level
expected for secondary N production, suggesting a delay in the release of the secondary N or a tertiary N production mechanism. We constrain the electron density using Si~II$^{*}$ and
C~II$^{*}$ absorption. We also report different metallicity vs. $\Delta V_{90}$ relations for sub-DLAs and DLAs. For two sub-DLAs with detections of emission lines from the underlying
galaxies, our measurements of the absorption-line metallicities are consistent with the emission-line metallicities, suggesting that metallicity gradients are not significant in these
galaxies.

\end{abstract}


\section{Introduction}

A probe of galaxy evolution, independent of galaxy flux, is provided by absorption lines in quasar spectra superposed by foreground galaxies along the sightlines, particularly the damped
Lyman-$\alpha$ (DLA; neutral hydrogen column densities $N_{\rm HI} \ge 2 \times 10^{20}$ cm$^{-2}$) and sub-DLA ($10^{19} \le N_{\rm HI} < 2 \times 10^{20}$ cm$^{-2}$; P\'eroux et al. 2003)
absorbers. DLAs and sub-DLAs are the primary neutral gas reservoir available for star formation (e.g., Storrie-Lombardi \& Wolfe 2000; P\'eroux et al. 2005; Prochaska et al. 2005; Noterdaeme
et al. 2009, 2012; Zafar et al. 2013). Furthermore, they offer the most precise element abundance measurements in distant galaxies, independent of electron temperature, complementing the
emission-line abundances in Lyman-break galaxies (LBG's) or star-forming UV selected galaxies (e.g., Shapley et al. 2004; Erb et al. 2006). The metallicities, relative abundances, and gas
kinematics of DLAs/sub-DLAs provide unique indicators of the star formation history, regardless of galaxy redshift or morphology. In the era of precision cosmology, the snapshots of galaxy
properties offered by DLAs/sub-DLAs are crucial for constraining the co-moving densities of gas and metals and their evolution (e.g., Kulkarni et al. 2007).

Nearly undepleted elements such as Zn and S in DLAs/sub-DLAs offer direct ``dust-free'' metallicity indicators. Measurements of Zn or S abundances exist for many DLAs. Like DLAs, sub-DLAs
also show damping wings in the H I Ly-$\alpha$ absorption line profiles and hence allow reliable determinations of the H I column densities. The ionization corrections for sub-DLAs are
relatively modest, compared to those for the Lyman Limit Systems (LLS) (e.g., Dessauges-Zavadsky et al. 2003; Meiring et al. 2007, 2009b). Sub-DLAs are thus also well-suited for measuring
element abundances. However, until recently, most element abundance studies largely ignored sub-DLAs due to their lower $N_{\rm H I}$ and the need for high-resolution spectrographs. Further,
most earlier studies ignored absorbers at redshifts $z < 1.6$ (corresponding to $\sim 70 \%$ of cosmic history) due to the need for ultraviolet spectra to access the H I Ly-$\alpha$ line (for
measuring $N_{\rm{HI}}$) and S II $\lambda \lambda \, 1250, 1253, 1259$ (for measuring $N_{\rm{S II}}$) at $z < 1.6$, and Zn II $\lambda \lambda \, 2026, 2062$ (for measuring $N_{\rm{ZnII}}$)
at $z <  0.6$.  

Most chemical evolution models predict that the mass-weighted mean metallicity of galaxies should rise from $\la 0.1$ solar at $z > 3$ to $\approx$ solar at $z=0$ (e.g., Pei et al. 1999;
Somerville et al. 2001; Cora et al. 2003). Surprisingly, recent studies show that the DLA global mean metallicity is only $\sim 0.2$ solar even at low redshifts, and does not reach the
near-solar value predicted for the global mean metallicity at $z=0$ (e.g., Kulkarni et al. 2005, 2007, 2010; P\'eroux et al. 2006b; and references therein; Battisti et al. 2012). A missing
metals problem was also noted in high-$z$ DLAs from star formation rate (SFR) estimates  based on C II* absorption (Wolfe et al. 2003). In fact, after adding up the metal content of known
DLAs, Ly-$\alpha$ forest, UV-selected galaxies, and sub-mm galaxies, $\ga 1/3$ of the predicted metals at $z \sim 2$ appear to be missing (e.g., Pettini 2006; Bouche et al. 2007; but see also
\citealt{Peeps14}).

Interestingly, recent ground-based studies by our group and others have uncovered several near-solar or super-solar metallicity sub-DLAs at $0.6 < z < 1.5$ (P\'eroux et al. 2006a, 2008;
Meiring et al. 2007, 2008, 2009a, 2009b; Nestor et al. 2008; Dessauges-Zavadsky et al. 2009; and references therein). Indeed, $\sim 35 \%$ of sub-DLAs with Zn data at $0.6 < z < 1$ have solar
or super-solar metallicity. Moreover, supersolar sub-DLAs appear to exist even at $1.8 \la z \la 2.1$ (Prochaska et al. 2006; Som et al. 2013).  An extrapolation of the $z > 0.6$ data to
lower redshifts suggests that the mean metallicity of sub-DLAs is solar at $z \sim 0$  (e.g., Kulkarni et al. 2007, 2010; see also York et al. 2006, Prochaska et al. 2006). In fact, sub-DLAs
may contribute several times more than DLAs to the cosmic metal budget at $z < 1$ (\citealt{Kh07,Kul07}).  

Despite the above-mentioned developments, several issues remain unanswered regarding the chemical evolution of sub-DLAs. One of the main issues is whether the observed trends are reliable,
given the small-number statistics, especially at $z < 0.6$, where the samples are particularly sparse. Second, the role of ionization needs to be addressed directly for low-$z$ sub-DLAs. The
kinematics of low-$z$ sub-DLAs also has yet to be investigated. Above all, the relationship of the DLA and sub-DLA populations remains unclear. With the goal of improving this situation, we
carried out a mini-survey of sub-DLAs at $z < 0.5$ using the UV wavelength coverage offered by the {\it Hubble Space Telescope} (HST) Comic Origins Spectrograph (COS). Here we report the
results of this survey.

Section 2 summarizes the observations and data reduction. Section 3 describes the results obtained for each absorber. Section 4 examines the metallicity evolution of sub-DLAs in light of the
data presented here, and discusses the implications of our study for galaxy chemical evolution. Section 5 summarizes our conclusions. Throughout this paper, we assume the concordance
cosmology model with $\Omega_{m} = 0.3$, $\Omega_{\Lambda} = 0.7$, and $H_{0} = 70$ km s$^{-1}$ Mpc$^{-1}$.

\section{OBSERVATIONS AND DATA ANALYSIS}

\subsection{Sample Selection and Observations}

We observed four sub-DLAs at $0.1 < z < 0.5$ toward quasars with reasonably high NUV/FUV fluxes from GALEX (NUV mag =15.70-17.77 and FUV mag =16.36-18.34). Table 1 lists our targets. These
absorbers were chosen to have 19.0 $\le$ log $N_{\rm HI} <$  20.3, as determined from archival low-resolution UV spectra from HST FOS/STIS (e.g., Rao et al. 2006) and there was no further
selection based on strong metal lines. Also listed in Table 1 are the spectroscopic redshifts of candidate absorbing galaxies, if known, and the impact parameters of the quasar sightlines
from the centers of these galaxies.

The COS spectra presented here were acquired as part of the HST program GO 12536. The observations were obtained with the COS FUV or NUV channels in the TIME-TAG mode with the G130M or G185M
grating. Each sightline was observed in multiple exposures spanning four offset positions (FP-POS=1-4) for each central wavelength setting in order to minimize the effects of fixed-pattern
noise in the detector. For PHL1226 and PHL1598, one exposure each (for 1412 s and 1462 s, respectively) was lost due to an error in the fine guidance sensors. Table 2 summarizes the
observations. 

Our main goal was to measure the absorption lines of Fe II, Fe III, S II, S III, Si~II, and Si III. In addition, our settings covered lines of C II, C II$^{*}$, N I, N V, O I, O~VI, Si~II$^{*}$,
P II, Ar I, Mn II, and/or Ni II for some absorbers. The presence of multiple lines for most ions helps to remove confusion with Ly-$\alpha$ forest lines and to correct for saturation effects.
The exposure times were designed using the COS online Exposure Time Calculator and the FUV/NUV fluxes for our target quasars. Our goal was to reach a 3$\sigma$ sensitivity of 0.1 solar
metallicity in our absorbers. Using the H I column density of each absorber, we estimated the S~II column density corresponding to a 0.1 solar metallicity, and hence the corresponding
equivalent width limit needed to detect the S II $\lambda 1253$ line at a 3 $\sigma$ level. We designed the exposure times so as to reach the corresponding signal-to-noise ratio (S/N) needed
for each object to reach this limit.  
 
\subsection{Data Reduction}
 
The spectra were reduced and extracted with standard Image Reduction and Analysis Facility (IRAF) and Space Telescope Science Data Analysis System (STSDAS) packages. The data were processed
with CALCOS (version 2.19.7) during retrieval via the ``On the Fly Reprocessing" (OTFR) system. For PHL 1226 and PHL 1598, the OTFR-processing included the compromised exposures mentioned
above; therefore these OTFR-processed files were not used in further analysis. The raw data for these two sightlines, after retrieval, were reprocessed without the compromised exposures using
CALCOS. The pipeline processed raw TIME-TAG data from each exposure into a flat-fielded, background-subtracted, flux- and wavelength-calibrated one-dimensional extracted (x1d) spectrum. (See 
Massa \& York 2013 for a detailed description of the data flow through FUV and NUV TIME-TAG spectroscopic pipelines.) The x1d files corresponding to all the exposures from a single visit were
then co-added by the pipeline to produce a single one-dimensional spectrum. The spectra from multiple visits for a sightline (where applicable) were then combined using IRAF. The un-binned
spectra used for the analysis have dispersions of $\sim$ 9.6 m{\AA} pixel$^{-1}$ and $\sim$ 37 m{\AA} pixel$^{-1}$ for FUV and NUV, respectively. The S/N near the S II $\lambda\lambda$ 1250,
1253, 1259 triplet are in the range $\sim$ 7-8 pixel$^{-1}$ for the FUV data and $\sim$ 17-30 pixel$^{-1}$ for the NUV data. The spectra from different segments of the two detectors were
continuum-fitted using the IRAF ``CONTINUUM" task. For all spectra from a given detector segment, we experimented with continuum fitting using both cubic spline and Legendre polynomial
functions, and adopted the better fitting function (judged from the RMS of the residuals) for the continuum fit. The continuum-normalized spectra were then used for identifying and measuring
absorption features at the known sub-DLA redshifts.

\subsection{Equivalent Width Measurements and Column Density Determinations}
 
The equivalent widths of the absorption lines were measured using the package ``SPECP'' (developed by D. Welty and J. Lauroesch). The rest-frame equivalent widths derived from these
measurements are listed in Table 3. The 1$\sigma$ errors for the equivalent widths are given as well and include the effect of the photon noise and the uncertainty in continuum placement.
Cells marked with an ellipsis represent lines which could not be measured due to one or more of the following: lack of coverage, blending with Ly$\alpha$ forest lines, very poor S/N due to
spectrograph inefficiency at wavelength extremes, or coincidence of the line with artifacts.

Column density determinations were performed by fitting multi-component Voigt profiles to the observed absorption lines, using the programs VPFIT\footnote{http://www.ast.cam.ac.uk/~rfc/vpfit.html}
(version 10.0) and FITS6P \citep{Wel91}. These programs iteratively minimize the $\chi^{2}$ residual between the data and the theoretical Voigt profiles convolved with the instrumental
profile. Our profile fits incorporate the wavelength-dependent line spread functions of COS. A discussion of our profile-fitting technique can also be found in Khare et al. (2004). Estimation
of the integrated column densities from the absorption profiles was done also via the apparent optical depth (AOD; see Savage \& Sembach 1991) method using the program ``SPECP". Determination
of integrated column density using the AOD method is independent of the velocity structure model and provides a check of the total column densities derived using the multi-component profile
fits. The atomic data for all lines were adopted from Morton (2003). While a few oscillator strengths have been improved recently (see, e.g., Kisielius et al. 2014 for S~II), the changes are
relatively small (e.g., $\approx 0.04$ dex for S~II).

For the absorbers toward PHL 1226 and PKS 0439-433, our data covered the H I  Ly-$\alpha$ absorption lines. In these cases, the H I column densities were estimated using Voigt profile fitting.
These Voigt profile fits are shown in Fig. 1. Our $N_{\rm H I}$ values match within $1-2 \sigma$ with those estimated from lower resolution HST spectra in past studies (e.g., Rao et al. 2006).
The Voigt profile fits to the metal absorption line profiles for the absorbers required multiple components. The moderate spectral resolution and S/N of our COS spectra limited the number of
components we can resolve. For each sub-DLA, we adopted the minimum number of velocity components that were necessary to explain all the absorption lines observed. A velocity component for an
ion species was included in our fits only if it was consistent with at least two absorption line profiles, but in general, all the velocity components considered were present in most strong
line profiles.

The effective Doppler parameter ($b_{eff}$) and velocity ($v$) values of the components were determined by fitting multiple lines simultaneously. For the strong components, $b_{eff}$ and $v$
were determined from weaker and less saturated lines, while for the weaker components, the stronger transitions were used. For each component, $b_{eff}$ and $v$ values were adjusted
iteratively until a consistent set of these values, providing the best fit to that component in the different lines, was found. The uncertainties in the $b_{eff}$ values were $\sim 10-20\%$,
and the uncertainties in the $v$ values were consistent with the spectral resolution. Absorption profiles of neutral (O~I, N~I, Ar~I), singly ionized (such as S~II, Si~II, Fe~II, P~II, Mn~II,
Ni~II, C~II) and doubly ionized (Fe~III, Si~III, S~III) metals in a system were fitted using the same set of $b_{eff}$ and $v$ values, because these ions showed absorption at common
velocities. In a few cases, some doubly ionized metals also showed absorption at other velocities, which was fitted with additional velocity components. The Voigt profile fits, obtained using
these $b_{eff}$ and $v$ values, were used to determine the column densities of various ions in individual components. If a multiplet was observed (such as S II $\lambda\lambda$ 1250, 1253,
1259; Si II $\lambda\lambda$ 1021, 1193, 1260; N I $\lambda\lambda$ 1199.6, 1200.2, 1200.7) the lines were fitted simultaneously. The contribution from Si II, as obtained from the typically
saturated Si II $\lambda$ 1193 or Si II $\lambda$ 1260 (if the former was not available), to the blend between Si II $\lambda$ 1190.2 and S III $\lambda$ 1190.4 was used to obtain an upper
limit on the contribution from S III. Absorption lines from N~V ($\lambda\lambda$ 1239, 1243) and O~VI ($\lambda\lambda$ 1032, 1038), detected in two absorbers in our sample, appear to be
wider and show less sub-component structure in comparison with the lower-ion line profiles. Consequently, the velocity structures adopted to fit the N~V and O~VI absorption lines were
independent of those adopted for the lower ions.

Figs. 2-7 show the velocity plots for the observed metal lines in the various absorbers in our sample. Tables 4-9 list the velocities with respect to the absorber redshifts reported in
the literature (based on measurements of the Ly-$\alpha$ line), Doppler parameters, and column densities for each component fitted to the various metal ions for the absorbers. The column
densities in the weaker components that could not be well-constrained due to noise are each marked with an ellipsis in Tables 4-9; their contributions to the total column densities are
negligible. The column densities of individual components were added to derive total column densities, and the uncertainties in the total column densities were estimated by adding in
quadrature the uncertainties in the column densities of the individual components. The total column densities thus determined were compared with the integrated column densities estimated
using the AOD method and were found to agree closely, as can be seen from Tables 10-13. The uncertainties in the total column densities derived from the profile fits were comparable to
the uncertainties in the integrated column densities from the AOD method (which include the effect of photon noise and the uncertainty in continuum placement). In case a line was not
detected, the limiting equivalent width was determined from the local S/N, and a corresponding 3$\sigma$ column density upper limit was determined, assuming a linear curve of growth. For
a saturated line, the total column density measurement using the AOD method was adopted as the column density lower limit.

We note that, at the spectral resolution and S/N of our COS data, it is difficult to determine the velocity structures of the absorbers with high degree of accuracy. The velocity
structure models adopted in our Voigt profile fits are thus approximate and are not meant to give exact kinematic information about the absorbing gas. However, as mentioned before, the total
column densities derived from our profile fits agree well, within the uncertainties, with the integrated column densities derived via the AOD method which is independent of any velocity
structure model. We stress that our main goal is to determine the total element abundances (combined over all velocity components), for which the integrated column densities inferred from our
fits as well as AOD calculations are adequate.

\section{RESULTS}

Element abundances were estimated for each absorption system using the total metal column densities (summed over all the velocity components) derived from our data and the H I column
densities (from our data if available, or using values published in the literature). Throughout this paper, we use the standard notation [X/Y] = log $(X/Y)$ - log~$(X/Y)_{\odot}$ to denote
the relative abundances of elements X and Y. The reference solar abundances log $(X/Y)_{\odot}$ were adopted from the photospheric values of \citet{Asp09}. Table 14 lists the element
abundances derived for each of our sub-DLAs. These estimates do not include ionization corrections; but we examine ionization corrections in detail in section 3.3 and find them to be
relatively modest, typically $\la 0.2$ dex.

\subsection{Notes for Individual Objects}

\subsubsection{PHL 1226, $z_{abs}= 0.1602$}

Bergeron et al. (1988) discovered two spiral galaxies of Sbc and Scd types at $z=0.1592$ and $z=0.1597$ with angular separations of 6.4$\arcsec$ and 10.9$\arcsec$  from the quasar PHL 1226,
corresponding to impact parameters of 17.7 and 30.1 kpc, respectively. They also discovered an Mg~II absorption system at $z=0.1602$ in the spectrum of PHL 1226, and suggested that this
absorber was associated with either of the two spiral galaxies.

From an analysis of archival HST FOS spectra, Rao \& Turnshek (2000) found that the $z=0.1602$ system is not a DLA. Analysis of these archival HST spectra suggested an H I column density of
log $N_{\rm H I} = 19.70^{+0.15}_{-0.22}$ (see Christensen et al. 2005). Our HST COS observations gave a higher quality spectrum of the Ly-$\alpha$ line. The left panel of Fig. 1 shows the
result of Voigt profile fitting for this Ly-$\alpha$ line, from which we obtain log $N_{\rm  H I} = 19.48 \pm 0.10$. Our COS data also targeted lines of S II, S III, Si II, Si III, Fe II,
C II, P II, N I, N V, O I, and O VI. Figs. 2 and 3 show the velocity plots for the various metal lines in this absorber. Tables 4 and 5 list the results of the Voigt profile fits for the
lower ions and the higher ions, respectively. Table 10 summarizes the total column densities derived from our data.

Strong N II, N V, and O VI absorption is detected in this system. The components in the low ions show a velocity spread of $\sim 200$ km s$^{-1}$, while the higher ions extend over a total
of $\sim 460$ km s$^{-1}$. The O VI $\lambda 1032$ profile is especially strong and striking, and shows strong asymmetry, possibly suggesting the presence of outflows. The O VI $\lambda 1037$
line is consistent with O VI $\lambda 1032$, but has additional absorption in some components due to blends with other unrelated absorption. 

\subsubsection{PKS 0439-433, $z_{abs} = 0.1012$}

This absorber is known to have strong absorption lines of Mg II, Fe II, Si II, Al II, and C IV. A galaxy at $z=0.101$ has been detected at an impact parameter of 7.6 kpc from the quasar
(Petitjean et al. 1996; Chen et al. 2005). Petitjean et al. (1996) suggested that the Mg II absorber in this sightline is a DLA. Kanekar et al. (2001) reported a weak 21-cm absorber in this
sightline, with an estimated spin temperature of $\ga 730$ K. Based on an absence of 21-cm emission, they estimate a 3$\sigma$ upper limit to the gas mass of the absorber of $2.25 \times
10^{9} M_{\odot}$. Thus, they claim that this galaxy is not a large, gas-rich spiral. 

The right panel of Fig. 1 shows our Voigt profile fit to the Ly-$\alpha$ feature in this absorber, which gives the H I column density of log $N_{\rm H I} = 19.63 \pm 0.15$. Our COS spectra of
this absorber reveal lines of several metal ions, i.e. N I, N II, N V, O~VI, Si II, Si II*, P II, S II, S III, Ar~I, Mn II, Fe~II, and Fe III. Figs. 4 and 5 show the velocity plots for the
various metal lines in this absorber. Table 6 lists the Voigt profile fitting results for the lower ions, while Table 7 lists the results for the higher ions. Table 11 summarizes the total
column densities.

This absorber shows strong N II absorption and also shows N V and O VI. This indicates a significant amount of ionized gas in this absorber. The velocity spread of the higher and lower ions
is comparable, and relatively large ($\sim 400$ km s$^{-1}$). We note that the O VI column densities in this absorber as well as in the sub-DLA toward PHL 1226 are found to be much higher
than in diffuse interstellar clouds in the Milky Way, such as that toward $\alpha$~Vir (e.g., \citealt{YK79}). However, the $N_{\rm O VI}$ values for these low-$z$ sub-DLAs are comparable to
those reported by \citet{Leh14} for sub-DLAs at $z \sim 2-3$.

\subsubsection{TXS 0454-220, $z_{abs} = 0.4744$}

Rao et al. (2006) report log $N_{\rm H I} = 19.45^{+0.02}_{-0.03}$ for this absorber, based on archival HST UV spectra. Robertson et al. (1988) reported the discovery of Mg II, Ca II, and
Fe II absorption in this absorber from moderate-resolution spectra obtained at the Anglo-Australian Telescope (AAT). Based on curve of growth analysis, they reported  log $N_{\rm Mg II} =
15.48$, log $N_{\rm Ca II} = 12.20$, and log $N_{\rm Fe II} = 14.90$. Based on data from the Keck High-Resolution Echelle Spectrometer (HIRES), several velocity components appear to be
present in this system (Churchill et al. 2000). Churchill \& Vogt (2001) provide column density estimates for Mg I, Mg II, and Fe II from the HIRES data. Using the apparent optical depth (AOD)
method, they estimate total column densities log $N_{\rm Mg II}  > 14.27$ and log $N_{\rm Fe II} > 14.52$, which are consistent with the estimates of Robertson et al. (1988). For Mg I,
Robertson et al. (1988) estimated  log $N_{\rm Mg I} = 12.18$, while Churchill \& Vogt (2001) reported log $N_{\rm Mg I}$ = $12.53 \pm 0.01$. The latter value is likely to be more accurate
than the former, owing to the much higher spectral resolution of the Keck data (0.09 {\AA}) compared to that of the AAT data (1.4-2.0 {\AA}); in any case, both Mg I values are far smaller
than the Mg II column density. These measurements (without ionization corrections) would imply [Mg/H] = 0.43, [Fe/H] = -0.05, and [Ca/H] = -1.59 dex. The relative abundances [Ca/Fe] = -1.54
and [Ca/Mg] = -2.02 imply large depletions of Ca. 
 
Fig. 6 shows velocity plots from our HST COS data for the various metal lines in this absorber. Table 8 lists the Voigt profile fitting results. Table 12 summarizes the total column densities.
Our measurements indicate [S/H] = $0.47 \pm 0.04$, in close agreement with the [Mg/H] value estimated from the data of Robertson et al. (1988). Likewise, our measurement of [Ni/H] = $0.02
\pm 0.08$ is in close agreement with [Fe/H] = -0.05 estimated by Robertson et al. (1988). This agreement is reassuring, given that S and Mg are both $\alpha$ elements, while Ni and Fe are
both Fe-group elements. Nucleosynthetic differences are not expected between S and Mg, or between Fe and Ni. [Mg/S] = -0.04 would suggest that dust depletion may not play a significant role
in this absorber. Moreover, this absorber appears to show an intrinsic nucleosynthetic enhancement of the $\alpha$-elements with respect to the Fe-group elements.
 
The components detected in the low ions covered in our data show a velocity spread of $\sim 170$ km s$^{-1}$. Our data did not cover N II $\lambda 1084$, or lines of higher ions such as N V
$\lambda \, \lambda 1239, 1243$, O VI $\lambda \, \lambda 1032, 1037$ or Si III $\lambda 1207$ in this absorber.
 
\subsubsection{PHL 1598, $z_{abs} = 0.4297$}

This system was originally reported by Weymann et al. (1979). Bergeron (1986) discovered a galaxy at a redshift 0.430, located 8.6$\arcsec$ away from the quasar, corresponding to an impact
parameter of 48.2 kpc. Rao et al. (2006) reported log $N_{\rm H I} = 19.18 \pm 0.03 $ for this system based on archival HST UV spectra. 

Fig. 7 shows velocity plots from our COS data for the various metal lines in this absorber. Table 9 lists the Voigt profile fitting results. Table 13 summarizes the total column densities. We
detect S II, Si II, and C II clearly in this absorber, with mutually consistent velocity structures. We also constrain the column density of S III, based on the $\lambda 1190.2$ line. Owing
to the blend with Si II $\lambda 1190.4$, we can only put an upper limit on the S~III column density.

Based on Mg II lines detected in Keck High Resolution Echelle Spectrograph (HIRES) data for this absorber, Churchill \& Vogt (2001) reported log $N_{\rm Mg II} = 13.47 \pm 0.003$ using the
AOD method. However, this column density should be treated as a lower limit given that the Mg II $\lambda\lambda$ 2796, 2803 lines in this system are saturated. We retrieved archival HIRES
data (Program ID: C99H; PI: C. Steidel) for this sightline which had higher S/N compared to the data used in Churchill \& Vogt (2001) and resampled the data to match the spectral resolution
of our COS-NUV data (see the bottom right panel in Fig. \ref{fig7}). While a perfect agreement between the velocity structures inferred from HST COS and Keck HIRES data is not expected due to
differences in S/N and wavelength calibration, the Mg II $\lambda\lambda$ 2796, 2803 line profiles are reasonably well-fitted by the velocity structure described in Table \ref{phl1598velstruc}.
The Voigt profile fit and the AOD measurement yield log $N_{\rm Mg II} > 13.83$ and log $N_{\rm Mg II} > 13.50$, respectively, from the Mg II $\lambda$ 2796 line. The corresponding values
determined from the relatively less saturated Mg II $\lambda$ 2803 line are log $N_{\rm Mg II} > 13.90$ and log $N_{\rm Mg II} > 13.71$.

Our results indicate that the S abundance is near-solar in this absorber, while both Si and C are considerably sub-solar. This cannot entirely be attributed to saturation effects as S II
$\lambda \, \lambda 1253, 1259$ and Si II $\lambda \, 1190$ appear to be relatively unsaturated while Si II $\lambda \, 1260$ and C II $\lambda 1334$ appear to be only moderately saturated.
The low Si and C abundances do not appear to be a result of ionization effects either (see section 3.3). The underabundance of Si and C could be partly due to dust depletion. Indeed, [Mg/H]
$>-1.1$ dex, deduced from our AOD measurement for the Mg II $\lambda \, 2803$ line, also seems to indicate some dust depletion in this absorber.

HST FOS spectra showed the presence of relatively weak C IV and possibly Si IV absorption in this absorber (Churchill et al. 2000). Our data did not cover these lines, N II $\lambda 1084$,
Si III $\lambda 1207$, or O VI $\lambda \, \lambda 1032, 1037$. Our data did cover N V $\lambda\lambda 1239, 1243$, but these lines were not detected. Overall, this absorber appears to be
relatively modestly ionized. The components in the ions detected in our data (C II, Si II, S II) show a spread of $\sim 100$ km s$^{-1}$. Compared to the more strongly ionized absorbers
toward PHL 1226 and PKS 0439-433, this absorber appears to have a narrower velocity spread in the low ions.

\subsection{Abundance Patterns}

In all of our target sub-DLAs, the measured abundance of the primary metallicity indicator S was found to be near-solar or super-solar. In all absorbers except that toward PHL 1598, the S
abundance values are consistent with the constraints from Si, C, and/or O. In the absorber toward PHL 1598, the abundances of both C and Si are considerably sub-solar compared to S. This does
not appear to be an ionization or saturation effect, and may suggest significant dust depletion in this absorber. In the absorbers toward PHL 1226 and PKS 0439-433, Fe is mildly depleted with
respect to S, suggestive of modest dust depletion or $\alpha$/Fe enhancement. In the absorber toward TXS 0454-220, combining our results with Mg and Fe abundances from the literature suggests
that dust depletion is not important, and that there may be nucleosynthetic $\alpha$/Fe enhancement.

Since P is also a weakly depleted element, it is interesting to compare the S and P abundances. We measure [S/P] = $-0.30 \pm 0.19$ in the absorber toward PHL 1226 and $-0.04 \pm 0.07$ in the
absorber toward PKS 0439-433. The S and P abundances are thus consistent with being in the solar ratio. This is roughly similar to the finding of Battisti et al. (2012), who reported [S/P]
$=0.19 \pm 0.16$ and $0.19 \pm 0.21$ for two other sub-DLAs at $z < 0.3$. Thus, our values do not suggest a profound odd-even effect between these two elements.

The measured abundances of N are substantially sub-solar. This appears to be caused by nucleosynthetic effects, as we discuss further in section 4.4. The measurement of Ar abundance from Ar I
(available in only 1 system) is also substantially sub-solar; but this appears to be caused by ionization, as described in section 3.3.2. On the other hand, Mn/H appears to be super-solar in
the $z=0.1012$ absorber toward PKS 0439-433. This system shows [Mn/Fe] = $0.80 \pm 0.15$, resembling the high [Mn/Fe] values seen in our ground-based studies of higher redshift metal-rich
sub-DLAs (e.g., Meiring et al. 2009a; Som et al. 2013). As noted in those studies, higher redshift sub-DLAs appear to follow a [Mn/Fe] vs. metallicity trend that is steeper than that for
DLAs, and for abundances in the Milky Way and the Small Magellanic Cloud. In fact, the [Mn/Fe] value for the sub-DLA toward PKS 0439-433 lies substantially above the [Mn/Fe] vs. metallicity
trend for the higher redshift sub-DLAs. 

Finally, we note that while our abundance analysis is based on integrated column densities (summed over all velocity components), there appear to be some variations between different
components. For example, the S II and P II profiles in the absorbers toward both PHL 1226 and PKS 0439-433 seem to differ in different components, suggesting a variation of the S/P ratio
between the components. The Mn II and Fe II profiles in the absorber toward PKS 0439-433 also show differences between different components. A variation between components is also seen in the
N I / N II ratio in the absorbers toward PHL 1226 and PKS 0439-433, suggesting a variation in the ionizing radiation field and/or density. The  N I/ O I ratio in the absorber toward PHL 1226
also appears to vary between different components, which could suggest nucleosynthetic variations owing to varying stellar populations. The spectral resolution of our COS data is not
sufficient to meaningfully address these component-to-component variations. Observations at higher spectral resolution are essential to address such variations in detail.

\subsection{Ionization Effects}

In both DLA and sub-DLA  studies, abundances are often calculated from a limited number of ions that are observationally accessible. For example, since the dominant ionization stages of Zn
and S are usually Zn II and S II, [Zn/H] and [S/H] are often estimated from Zn~II/ H I and S~II/ H I, respectively. In principle, the true abundances could be different, depending on the
column densities of the higher ions (e.g. Zn III, S~III, H~II). It is, therefore, important to estimate the extent of these ionization corrections, defined as $\epsilon (X) = [X/H]_{total} -
[X_{dominant \, ion}/H I]$. 

The element abundances in DLAs are generally believed to be not much affected by ionization effects, since DLAs, due to their high $N_{\rm H I}$, are expected to cause self-shielding of
photons with $h\nu>13.6$ eV, i.e. photons capable of ionizing H I. On the other hand, Lyman limit systems and (especially) the Lyman-alpha forest are highly ionized owing to their lower H I
column densities. Given that the H I column densities of sub-DLAs lie in between those of DLAs and Lyman-limit systems, sub-DLAs are likely to contain some ionized gas. (Indeed, as described
earlier, strong N II, N V, and O VI absorption is observed toward two of our sub-DLAs.) This raises the question of whether ionization corrections could be the possible explanation of the
high metallicities seen in higher-redshift sub-DLAs. However, several ionization modeling studies of sub-DLAs at $z \ga 0.6$ showed that while these sub-DLAs can be significantly ionized,
ionization corrections cannot explain their high element abundances; in fact, the ionization correction for Zn is positive, i.e. the true [Zn/H] is even {\em higher} than that estimated from
[Zn II/H I] (see, e.g., Meiring et al. 2007, 2009b; Som et al. 2013). 

We now inspect whether ionization corrections are significant in the $z < 0.5$ sub-DLAs. Given that the intergalactic UV radiation field is much less intense at low redshifts (e.g., Kulkarni
\& Fall 1993; Scott et al. 2002), the low-redshift sub-DLAs may be expected to be less ionized than their higher-redshift counterparts. On the other hand, if SFRs are especially high in these
low-$z$ sub-DLAs, the local radiation field could be stronger. 
 
To investigate the extent of ionization corrections, we performed photoionization model calculations for all the absorbers in our sample using version 13.01 of the plasma simulation code
{\em CLOUDY} (Ferland et al. 2013). The models assumed the ionizing radiation incident on the sub-DLA absorbing gas to be a combination of the intergalactic UV background (adopted from
\citealt{HM96} and \citealt{Mad99} and evaluated at the absorber redshift) and a radiation field produced by O/B type stars (based on a Kurucz model stellar spectrum for a temperature of
30,000 K), mixed in equal parts. The inclusion of the stellar component of the ionizing radiation in our models was motivated by suggestions that the contribution from local sources to the
ionization of quasar absorbers may not be negligible in comparison with the background ionizing radiation (e.g., \citealt{Sch06,BS14}). In addition to the radiation fields mentioned above,
our simulation includes the cosmic ray background as well as the cosmic microwave background at the absorber redshift. We note, however, that our models did not include radiation originating
in local shocks from white dwarfs, compact binary systems, or supernovae. For each absorber, grids of photoionization models, tailored to match the observed N$_{\rm H \ I}$ and the observed
metallicity based on N$_{\rm S II}$, were produced by varying the ionization parameter, defined as
$$ U=\frac{n_{\gamma}}{n_{H}}=\frac{\Phi_{912}}{cn_{H}}$$
(where $\Phi_{912}$ is the flux of radiation with h$\nu$ $>$ 13.6 eV), from $10^{-6}$ to 1. The simulated column density ratios between various ions were then compared with the observed
values to constrain the ionization parameter and estimate ionization corrections. In particular, we used column density ratios of the adjacent ions of the same element (e.g., Si III/ Si II, S
III/S II, or Fe III/Fe II), as they provide more reliable observational constraints than the ratios involving different elements, which may be affected by differential depletion or intrinsic
nucleosynthetic differences. We note that the shape of the ionizing spectrum strongly influences ionization in the gas and therefore the results of our simulations are sensitive to the
assumption for the incident spectrum. We also note that our models do not use the observed column densities of O VI, since O VI may be largely collisionally ionized. Indeed, in the absorbers
toward PHL 1226 and PKS 0439-433, the column density of O VI implied by our photo-ionization models is about 1/1000 of that observed; thus, most of the O VI in these absorbers appears to be
collisionally ionized.

We caution that our photoionization models are, as is usual in such models, based on several simplifying assumptions. The estimates of ionization corrections should therefore be treated as
indicative, since they are sensitive to various parameters such as the radiation field, cloud geometry, gas density, abundance pattern, etc. For example, the assumptions of uniform density
and radiation field for the different velocity components is likely be too simplistic, given the presence of multi-phase gas seen from a comparison of the higher and lower ions. However, to
obtain more realistic models fitting each component separately, it would be necessary to determine the H I column density in individual velocity components (e.g. from higher Lyman-series
absorption lines; see \citealt{Proc10}). The lack of information in the existing data regarding $N_{\rm H I}$ in individual components, together with the lack of knowledge of the spatial
positioning of the gas at various velocity components and their effects on radiative transfer, make it difficult to perform more sophisticated photoionization modeling. However, as we note in
section 4.3.1, the model results appear to be consistent with the observational constraint on electron density from a Si II fine structure line in at least one absorber where we have column
density estimates of Si II$^{*}$ and Si II.

We now describe the results of our photo-ionization simulations for the individual systems. These results are summarized in Fig. \ref{fig8} for the sub-DLAs toward PHL 1226 and PKS 0439-433,
and in Fig. \ref{fig9} for the sub-DLAs toward TXS 0454-220 and PHL 1598. 

\subsubsection{PHL 1226, $z_{abs}=0.1602$} 

For this system, the observed lower limit on the $N_{\rm Si III}$ to $N_{\rm Si II}$ ratio was used to obtain a lower limit on the ionization parameter of log $U >$ -2.75. The observed upper
limit on $N_{\rm S III}/N_{\rm S II}$ implies log U $<$ -2.45. These results suggest that the ionization correction for [S/H] ranges between -0.24 dex to -0.29 dex. We adopt the correction to
metallicity to be -0.26 dex derived for log $U = -2.60$ (the mean value of the ionization parameter range described above), which implies an ionization-corrected value of [S/H] =  -0.02. The
corresponding ionization correction for the N abundance derived from N I/ H I is +0.05 dex. For C, the ionization correction for the abundance derived from C II/ H I is -0.40 dex, i.e., the
ionization-corrected [C/H] is $> -1.0$ dex. 

\subsubsection{PKS 0439-433, $z_{abs}=0.1012$} 

The observed ratio of $N_{\rm Fe III}/N_{\rm Fe II}$ in this absorber suggests  log $U = -2.3$, implying a correction of -0.18 dex for [S/H], and therefore, an ionization-corrected [S/H] of
$0.1$ dex. The corresponding corrections for Ar and N are +0.91 dex and +0.05 dex. After making the large Ar ionization correction, [Ar/H] is $0.15$ dex, consistent within $\sim 1 \sigma$
with [S/H]. In other words, the low observed Ar I/H I in this system can be explained as an ionization effect.

\subsubsection{TXS 0454-220, $z_{abs}=0.4744$} 

For this system, S is the only element available in multiple ionization stages. The upper limit on $N_{\rm S III}$/ $N_{\rm S II}$ allowed us to place an upper limit on the ionization
parameter of log $U <$ -2.6. This implies that the ionization-corrected [S/H] could be lower by at most $0.20$ dex, i.e., [S/H] $> 0.29$ dex. For N, the correction is $\la 0.06$ dex. For C,
the ionization-corrected [C/H] could be lower by $ \la 0.24$ dex.

\subsubsection{PHL 1598, $z_{abs} = 0.4297$}

With log $N_{\rm H I}$ = 19.18, this sub-DLA has the lowest $N_{\rm H I}$ value in our sample. The observed upper limit on the $N_{\rm S III}$ to N$_{\rm S II}$ ratio implies an upper limit
on the ionization parameter of log $U$ $<$ -2.8, which, in turn, suggests that the observed [S II/ H I] could overestimate the true [S/H] by $< 0.46$ dex. We note, however, that the true
ionization correction is likely to be much smaller. This is because the S III $\lambda 1190.2$ line in this absorber is very noisy, and may be blended with not only Si II $\lambda 1190.4$,
but possibly also with Ly-$\alpha$ forest lines and/or noise. Thus, $N_{\rm S III}$ and hence the S ionization correction are likely to be smaller. The corresponding estimate of the C
ionization correction implies that the C abundance could be lower than that estimated from C II/H I by at most $ 0.5$ dex. This means that the much lower observed abundance of C compared to
S cannot be explained as an ionization effect.

\section{DISCUSSION} \label{}

We now discuss the implications of our results for various aspects of galactic chemical evolution. In section 4.1, we combine our results with those for other sub-DLAs and DLAs from the
literature, and discuss the trends in metallicity vs. redshift. We also discuss the effects of ionization in sub-DLAs on these trends. We discuss N abundances in section 4.2. We put
constraints on the electron density in the absorbing gas using Si~II$^{*}$ and C~II$^{*}$ absorption in section 4.3. In section 4.4, we present kinematic measurements for the sub-DLAs in our
HST study and examine the relation between metallicity and velocity dispersion for DLAs and sub-DLAs. Finally, in section 4.5, we compare the absorption-line metallicities along the
quasar sightlines from our data with emission-line metallicities near the centers of the foreground galaxies (available in the literature) for two of our sub-DLAs.

\subsection{Metallicity-Redshift Relation}

We now examine the metallicity-redshift relation for the absorbers, combining our data with Zn- or S-based metallicity measurements for sub-DLAs and DLAs from the literature (\citealt{Aker05,
Bat12,Boi98,Cent03,DLV00,DZ03,DZ09,EL01,Fyn11,Ge01,Kh04,Kul99,Kul05,Led06,Lop99,Lop02,LE03,Lu95,Lu96,Mei06,Mei07,Mei08,Mei09a,MY92,Mey95,Mol00,Nest08,Not08,Per02,Per06a,Per06b,Per08,Petit00,
Pet94,Pet97,Pet99,Pet00,PW98,PW99,Pro01,Pro02,Pro03a,Pro03c,Raf12,Rao05,Som13,SP01}). We converted the metallicity measurements compiled from the literature to metallicities relative to the
set of solar abundances used in this work (from \citealt{Asp09}). The sub-DLA and DLA samples used in our analysis contain 72 and 195 metallicity measurements, respectively. We emphasize that
we include only the systems with Zn or S measurements because these nearly undepleted elements provide the most direct gas-phase metallicity estimates. (Other weakly depleted elements are not
well-measured in ground-based spectra for most DLAs and sub-DLAs at $z \la 2$ due to the lack of multiple, unsaturated lines.) We exclude systems with measurements of only Fe or Si (e.g.,
many DLAs from Rafelski et al. 2012), since these elements are prone to depletion (see, e.g., \citealt{SS96,Draine03,Jenk09}).

The observed metallicity vs. redshift data for the sub-DLA and DLA samples described above are shown in Fig. \ref{fig:scatter}. The metallicity measurements presented in this work,
constituting 80~$\%$ of the sub-DLA [S/H] measurements based on detected S lines and $\sim$~50~$\%$ of the entire sub-DLA [S/H] sample at $z \la 0.5$, are shown with green striped circles in
this figure.

To derive the H I column density-weighted mean metallicity vs. look-back time relations for the sub-DLA and DLA samples, the data were binned in redshift, with 11-13 systems per sub-DLA bin
and 16-17 systems per DLA bin. For the systems with no detection of Zn and S, upper limits on Zn were used and were treated with survival analysis. For each redshift bin, the $N_{\rm H
I}$-weighted mean metallicity and its 1$\sigma$ uncertainty (including both sampling and measurement uncertainties) were calculated using the procedures described in Kulkarni \& Fall (2002).
The results are shown in Fig. \ref{fig:bin}. The lowest redshift bin for sub-DLAs contains nine [S/H] measurements at $z \la 0.5$ (from this work and \citealt{Bat12}) and two [Zn/H]
measurements at $0.5 < z < 0.65$ (from \citealt{Pet00,Kh04}). The $N_{\rm H I}$-weighted mean metallicity for this bin is found to be high, in agreement with the extrapolation from higher
redshift sub-DLAs in past studies (e.g., Som et al. 2013). We note that our target absorbers were selected simply on the basis of their $N_{\rm H I}$ values, i.e., on the basis of being
identified as sub-DLAs at low redshifts based on HST UV surveys. While those UV surveys were designed to observe systems with Mg II $\lambda 2796$ lines of $\ge 0.5$ {\AA} rest-frame
equivalent widths, this does not pre-select sub-DLAs randomly chosen from this H~I sample toward higher metallicities. This is discussed in detail in \citet{Kul07,Kul10}. Indeed, no
correlation between metallicity and $W_{2796}$ is present in the sub-DLA sample used in our study. Furthermore, our 3 $\sigma$ [S/H] detection sensitivity of 0.1 solar is much lower than the
metallicities observed for all of the  sub-DLAs in our program. Thus, the high sub-DLA metallicities seen in our sample are not an artifact of sample selection or inadequate detection
sensitivity.

We performed linear regression fits to the binned $N_{\rm H I}$-weighted mean metallicity vs. median redshift relations for both DLAs and sub-DLAs. The slope of this fit gives a measure of
the rate of the global mean metallicity evolution, while the intercept is the predicted mean metallicity at zero redshift. For DLAs, the best-fitting relation has a slope of $-0.192 \pm
0.047$ and an intercept of $-0.701 \pm 0.114$. For sub-DLAs, the best-fitting relation has a slope of $-0.311 \pm 0.095$ and an intercept of $0.047 \pm 0.159$. Thus, both DLAs and sub-DLAs
show metallicity evolution at the 4.1$\sigma$ and 3.3$\sigma$ levels, respectively. Moreover, the rates of metallicity evolution of DLAs and sub-DLAs appear to be similar, differing only at
the 1.1$\sigma$ level. However, the mean metallicity of sub-DLAs at zero redshift is clearly higher than that for DLAs, since the intercepts of the two relations differ at the $3.8 \sigma$
level. We stress that, despite the small size of our HST sample of low-$z$ sub-DLAs, the addition of these data has improved the accuracy of the sub-DLA intercept determination by $\sim 45
\%$. Furthermore, the addition of our HST data has established the metallicity evolution of sub-DLAs more definitively, and has shown more clearly that sub-DLAs have higher mean metallicity
than DLAs at all redshifts studied. The dot-dashed curve in Fig. 11 shows the prediction for the mean interstellar metallicity in the Pei et al. (1999) model with the optimum fit for the
cosmic infrared background intensity. The dot-double-dashed curve shows the mean metallicity of cold interstellar gas in the semi-analytic model of Somerville et al. (2001). These models
appear to agree better with the sub-DLA data than with the DLA data.

We next considered the effect of ionization corrections on sub-DLA metallicities and on the metallicity-redshift relation. To do this, we repeated the calculations described above considering
the ionization-corrected metallicity measurements for the sub-DLAs presented here as well as for the sub-DLAs with ionization correction estimates available in the literature (see Table
\ref{iontable}). In the case where the ionization correction for a system was estimated to lie in a range, we adopted the mean correction value for our analysis. For two absorbers in our
sample for which we only have estimates of the maximum amount of ionization corrections, we adopted these maximum corrections to the observed metallicities, to be conservative. We note,
however, that the true ionization corrections for these systems are likely to be smaller. The ionization-corrected sub-DLA metallicity vs. redshift data are shown in Fig. \ref{fig:scatter},
where the corrected measurements and the corresponding observed values before corrections are connected using vertical brown lines.

In Fig. \ref{fig:bin}, we show the effect of ionization corrections on the $N_{\rm H I}$-weighted mean metallicity vs. redshift relation for sub-DLAs. The black squares in Fig. \ref{fig:bin}
show this relation calculated after including ionization corrections for our sub-DLAs as well as those from the literature that have published ionization corrections. While many of the
sub-DLAs from the literature have no ionization corrections available, many of the low-$z$ sub-DLAs do have ionization corrections. We note, in particular, that all sub-DLAs at $z \lesssim
0.5$ have ionization corrections available. The best-fitting linear regression fit to these $N_{\rm H I}$-weighted mean metallicity vs. redshift points gives a slope of $-0.183 \pm 0.116$ and
an intercept of $-0.103 \pm 0.185$. Thus, the difference between the sub-DLA and DLA mean metallicities at zero redshift appears to persist at the $\sim 2.8 \sigma$ level.

For the sake of completeness, and in an attempt to consider the effect of ionization corrections for all sub-DLAs, we next made approximate corrections to the S- and Zn-based metallicity
values for those sub-DLAs that do not have any published ionization corrections. To do this, we corrected the S-based metallicity values by -0.2 dex (the median value of the available
ionization corrections for S), and the Zn-based metallicity values by +0.15 dex (median of the available ionization corrections for Zn). The corresponding $N_{\rm H I}$-weighted mean
metallicity vs. redshift points are shown in Fig. \ref{fig:bin} as green squares. The best-fitting linear regression fit to these points gives a slope of $-0.206 \pm 0.118$ and an intercept
of $0.022 \pm 0.195$, which  implies a $3.3 \sigma$ difference between the zero-redshift mean metallicities of DLAs and sub-DLAs.

Finally, we note that it is not possible to obtain homogeneously ionization-corrected values by just compiling values from multiple studies, given the various uncertainties associated with
the radiation field, depletion pattern, etc. The only way to obtain a homogeneously calculated set of ionization corrections would be to run ionization models in a consistent manner on all
systems in the literature. This would be a much broader study that is beyond the scope of the present paper. However, based on the analysis presented here, the large difference between the
intercepts of the best-fitting metallicity-redshift relations derived above for DLAs and sub-DLAs is expected to persist, even after incorporating ionization corrections for all the absorbers
in the sample.

\subsection{Nitrogen Abundances}

The abundance of N relative to O or other $\alpha$-elements in DLAs/sub-DLAs is of great interest because it can provide insights into the nucleosynthetic production of N. It is believed that
N is produced through both a primary process from C synthesized in the star itself, and a secondary process involving C and O from previous star formation i.e., from C and O inherited by the
star from the interstellar medium that it formed out of (e.g., \citealt{EP78,RV81,VE93,Hen00}). Primary N is believed to be produced in mainly intermediate-mass stars, while secondary N
production can occur in stars of all masses. The $\alpha$-elements are believed to be produced predominantly in high-mass stars. The primary component is expected to dominate at low
metallicities, while the secondary component is expected to dominate at high metallicities. Indeed, the ratio [N/$\alpha$] observed in H II regions can be accounted for by primary N
production in nearby low-luminosity galaxies (e.g., Van Zee \& Hayes 2006) or blue compact dwarf galaxies (Izotov \& Thuan 2004), and by secondary N production in the more metal-rich nearby
spiral galaxies (e.g., Van Zee et al. 1998).

In all three sub-DLAs where our data covered lines of N, the measured abundance of N is found to be much lower than that of S, with [N/S] ranging between -1.83 and -0.79 dex. This finding is
similar to, but even more extreme than, that of Battisti et al. (2012), who found [N/S] between -1.12 and -0.58 dex for the sub-DLAs in their sample.

Fig. 12 shows the [N/$\alpha$] vs. [$\alpha$/H] data for sub-DLAs at $z < 0.5$ from our work and Battisti et al. (2012) and for sub-DLAs at $z > 2$ (Zafar et al. 2014 and references therein).
For clarity, we have plotted only the detections and excluded the limits. The $\alpha$ element used here is O or S. The solid symbols show the values without ionization corrections. The data
for $z < 0.5$ sub-DLAs appear quite distinct from those for sub-DLAs at $ z > 2$. The dashed lines denote approximate levels of primary N production and an extrapolation to lower metallicities
of the secondary N production trend observed at higher metallicities (see, e.g., \citealt{VE93}). If the release of the primary N component is delayed relative to the release of O, N/O values
between the pure secondary and primary + secondary levels are possible. If there is a delay in the release of the secondary component, then the N/O values can spread to the lower right across
the secondary line. Data for most DLAs and sub-DLAs at $z > 0.6$ are known to lie mainly between the two curves (e.g., \citealt{Pet95,Pet02,Cent03,Zaf14}). By contrast, the data for several
low-$z$ sub-DLAs lie to the lower right of the secondary N production line. 
 
Part of this deficit in [N/$\alpha$] could, in principle, arise from ionization effects. The ionization correction is more significant for S than for N, while it is negligible for O, as
expected from the first and second ionization potentials of S, N and O. To investigate how much effect ionization corrections have, we show with dashed unfilled symbols the values for the
$z < 0.5$ sub-DLAs including the maximum amount of ionization corrections allowed by the observed ion ratios. For the $z > 2$ sub-DLAs, ionization corrections are not available in the
literature.  It is clear that while ionization corrections change the exact values somewhat, several of the N detections in $z < 0.5$ sub-DLAs still lie below the secondary N production line.
This may suggest a different nucleosynthetic origin for N in these absorbers. Indeed, a tertiary N production mechanism has been suggested (e.g, \citealt{Hen00}). Alternately, the low
[N/$\alpha$] values may simply represent the lower end of the scatter around the mean trend expected for secondary N production, possibly arising due to a delay in the release of the
secondary N (e.g., \citealt{VE93}).

\subsection{Electron Density}

\subsubsection{Si II$^{*}$ Absorption}

The electron density of the absorbing gas can be estimated using fine structure absorption lines, for an assumed gas temperature (see e.g., \citealt{SP00}). The Si II upper level is expected
to be populated predominantly by collisional excitation, and depopulated predominantly by radiative de-excitation. Si II$^{*}$ absorption has been detected in gamma-ray burst afterglows (e.g.,
Savaglio et al. 2012 and references therein), but is rare in quasar absorbers. Kulkarni et al. (2012) reported the first Si II$^{*}$ absorption in an intervening quasar DLA. Here we report
possible presence of Si~II$^{*}$ $\lambda 1264.7$ absorption in the two low-$z$ sub-DLAs toward PHL 1226 and PKS 0439-433. In the sub-DLA at $z=0.1602$ toward PHL 1226, some Si II$^{*}$
$\lambda 1264.7$ absorption may be present at a $2.7 \sigma$ level in a component associated with strong absorption seen in the other low ions (see Fig. \ref{fig2}; Table \ref{ewtable}). The
Si~II$^{*} \, \lambda 1264.7$ line is detected at a higher significance level ($\sim 3.3 \sigma$) in the $z=0.1012$ sub-DLA toward PKS 0439-433 (see Fig. \ref{fig4}; Table \ref{ewtable}). The
corresponding Si II~$^{*}$ $\lambda 1194.5$ lines are, unfortunately, hard to deblend from the red components of the Si II $\lambda 1193.3$ lines. The Si II$^{*}$ $\lambda 1197.4$ lines in
these absorbers are located in noisy regions. However, we have checked that these noisy regions are consistent with the expected Si II$^{*}$ $\lambda 1197.4$ absorption based on the column
density of Si II$^{*}$ derived from the $\lambda 1264.7$ line.

In order to estimate the electron density, we assume equilibrium between collisional excitation and spontaneous radiative de-excitation. In terms of the collisional excitation rate $C_{12}$
and spontaneous radiative de-excitation rate $A_{21}$, the electron density is then given by $n_{e} = \bigl(  N_{\rm Si II^{*}}$/$N_{\rm Si II} \bigr) A_{21}/C_{12}$. We use the collisional
excitation rate for Si II $C_{12} = 3.32 \times 10^{-7}  \, (T/10,000)^{-0.5} \, exp(-413.4/T)$ cm$^{3}$ s$^{-1}$  and the spontaneous radiative de-excitation rate for Si~II$^{*}$ $A_{21} =
2.13 \times 10^{-4}$ s$^{-1}$ (Srianand \& Petitjean 2000). Given the temperature dependence of $C_{12}$, we consider two illustrative cases of $T= 500$ K and $T =7000$ K. 

For the $z=0.1602$ absorber toward PHL 1226, we find $N_{\rm Si II^{*}}$/$N_{\rm Si II} = 2.3 \times 10^{-3}$ and estimate the electron density to be $n_{e} = 0.77^{+0.48}_{-0.30}$ cm$^{-3}$
for $T=500$ K or $1.33^{+0.83}_{-0.51}$ cm$^{-3}$ for $T= 7000$ K. We compare this electron density to the density of H$^{+}$ found by our photoionization model for this absorber. Based on
the fraction of ionized H corresponding to the ionization parameter range suggested by our model, the ionized H density $n_{H +}$ is estimated to be in the range 0.73 to 0.86 cm$^{-3}$ with
the value of 0.79 cm$^{-3}$ at the mean log $U$ of the range. This value is in good agreement with the electron density $n_{e} = 0.77^{+0.48}_{-0.30}$ cm$^{-3}$ estimated above. It thus
appears that despite the various simplifying assumptions in our photoionization model, it describes several observational constraints reasonably well. It is also interesting to note that the
electron density in the sub-DLA toward PHL1226 is comparable to that in the $z=2.2$ super-DLA with log $N_{\rm H I} = 22.05$ toward J11350-0010 (0.53-0.91 cm$^{-3}$, Kulkarni et al. 2012). 

For the absorber toward PKS 0439-433, the Si~II$^{*} \, \lambda 1264.7$ line is well-detected, but the Si II $\lambda 1260$ line is saturated, allowing only an upper limit $N_{\rm Si
II^{*}}$/$N_{\rm Si II} <  2.4 \times 10^{-2}$. Using the upper limit on Si II$^{*}$/ Si II, we obtain only a weak upper limit on the electron density, i.e., $n_{e} < 8.0$ cm$^{-3}$ for
$T=500$ K or $n_{e}  <13.9$ cm$^{-3}$ for $T= 7000$ K. We note that the electron density in this absorber is likely to be much lower than this limit. This is because the Si II lines in this
latter absorber are heavily saturated; therefore the Si II column density is likely to be much higher, and the $N_{\rm Si II^{*}}$/$N_{\rm Si II}$ ratio is likely to be much lower. 

For the absorber toward TXS 0454-220, we place a 3$\sigma$ upper limit on the Si II$^{*}$ column density of log $N_{\rm Si II^{*}} < 11.86$ from the non-detection of the Si II$^{*}$ $\lambda$
1194.5 line. Combining this with the lower limit on the Si II column density of log $N_{\rm Si II} > 14.13$, we obtain an upper limit on the electron density of $ n_{e} < 1.6$ cm$^{-3}$ for
T=500 K or $ n_{e} < 2.7$ cm$^{-3}$ for T=7000 K.

\subsubsection{C II$^{*}$ Absorption}
The ratio of column densities of C II$^{*}$ and C II can also be used to constrain the electron density in the gas, for an assumed gas temperature, because the C II upper level is expected to
be populated by collisional excitation, and depopulated by radiative de-excitation. Equilibrium between these processes gives the relation $N_{\rm C II^{*}} /N_{\rm C II}= n_{e} C_{12}(T)/
A_{21}$, where $A_{21} = 2.29 \times 10^{-6}$ s$^{-1}$ (Nussbaumer \& Storey 1981). The collision rate coefficient is given by $C_{12}(T) = [8.63 \times 10^{-6} \Omega_{12} / (g_{1} T^{0.5})]
\, {\rm exp}(-E_{12}/kT)$ (Wood \& Linsky 1997), where $g_{1} = 2$, $E_{12} = 1.31 \times 10^{-14}$ erg, and the collision strength $\Omega_{12}$ depends on temperature only weakly. At
T$\sim 7000$ K, $\Omega_{12} = 2.81$, giving $C_{12} = 1.43 \times 10^{-7}$. 

For the absorber toward PHL 1226, it is not possible to constrain the C II$^{*}$ $\lambda 1037.0$ line since it lies in a noisy region and coincides in part with O VI $\lambda 1037.6$. For
the absorber toward PKS 0439-433, we cover neither C II nor C~II~$^{*}$.

For the absorber toward TXS 0454-220, from the non-detection of C II$^{*} \lambda 1335.7$, we put a 3$\sigma$ upper limit log $N_{\rm C II^{*}} < 12.57$. Combining this with the lower limit
log $N_{\rm C II} > 14.92$, we constrain the electron density to be $n_{e} < 2.9 \times 10^{-2}$ cm$^{-3}$ for $T = 500$ K or $n_{e} <5.9 \times 10^{-2}$ cm$^{-3}$ for $T = 7000$ K. This
constraint is more severe than the constraint from Si II$^{*}$ non-detection for this absorber, and suggests that this absorber is not highly ionized. 

For the absorber toward PHL 1598, we have a poorer 3 $\sigma$ upper limit log $N_{\rm C II^{*}}< 12.85$, as the C II$^{*} \lambda 1336$ line falls in a noisy region with an uncertain
continuum. Combining this with the measurement of log $N_{\rm  C II}$, we obtain the constraint on the electron density of $n_{e} < 2.4 \times 10^{-1}$ cm$^{-3}$ for $T = 500$ K or $n_{e}
< 4.9 \times 10^{-1}$ cm$^{-3}$ for $T = 7000$ K. 
 
\subsection{Kinematics}

The stellar mass and gas phase metallicity of star-forming galaxies have been found to be correlated at low as well as high redshifts (e.g., \citealt{Trem04,S05,Erb06}). A mass-metallicity
correlation has been suggested for DLAs as well (e.g., \citealt{Per03,Led06,Mol13}) using velocity width of absorption as a proxy for the mass of the absorber galaxy. The width of unsaturated,
low-ionization absorption line profiles in velocity space can be used to probe the kinematics of absorbing gas (see \citealt{WP98}). If the kinematics are primarily governed by the underlying
gravitational potential well, the velocity width determined from absorption lines can be considered as an indicator of mass. Here we present the velocity width values measured, following the
prescription of \citet{WP98}, for the low-$z$ sub-DLAs from our COS data and examine the velocity width versus metallicity (based only on weakly depleted elements) correlation for sub-DLAs
and DLAs. Table \ref{kinematicstable} lists the velocity width measurements from our COS data.

Figure \ref{Fig:kinematicsfig} shows the velocity width versus Zn- or S-based metallicity data for sub-DLAs and DLAs. The sub-DLA sample consists of the measurements presented here as well as
those from \cite{Led06}, \cite{Mei09b}, and \cite{Som13}. This sample contains 31 systems spanning a redshift range of $0.1 \la z \la 3.1$. The DLA sample, containing 71 systems at $0.1 \la z
\la 4.4$, is based on \citet{Led06}, and \cite{Mol13}. Again, we converted all the metallicity measurements compiled from the literature to use the same set of solar abundances (i.e., the
solar abundances from \citealt{Asp09}). The DLA data, in agreement with previous studies, appear to be well correlated and a Spearman rank-order correlation test on the sample reveals the
correlation co-efficient to be $\rho_s$ = 0.69 with a very low probability ($P(\rho_s)<0.001$) of no correlation. The Kendall's $\tau$ for this sample is 1.0 (with a $<0.001$ probability of
no correlation).

The sub-DLA data appear to be less correlated ($\rho_s=0.39$, $P(\rho_s)=0.03$; $\tau=0.59$, $P(\tau)=0.02$). Ionization corrections to the sub-DLA sample, made as described in section 4.1,
appear to have little effect on the correlation of the data as the corrected sample (also shown in Fig. \ref{Fig:kinematicsfig}) corresponds to $\rho_s=0.35$ with $P(\rho_s)=0.06$. An
inherent scatter in the DLA and sub-DLA data is indeed to be expected, as the observed velocity width includes effects of local velocities and turbulent motion in the gas on top of the
velocity dispersion due to the underlying gravitational potential well. Therefore, the large scatter in the sub-DLA data, although in part due to the smaller sample size, could indicate
higher levels of turbulence or large scale motion in these systems compared to DLAs. The sub-DLA and DLA data, considered together, appear to be well-correlated and a Spearman correlation
test on the combined sample yields $\rho_s$ = 0.55 with a $<0.001$ probability of no correlation. This suggests a general mass-metallicity relation exists among absorption-selected galaxies,
if the velocity dispersion offers a measure of the mass of the absorbing galaxy.

A linear regression fit to the sub-DLA data gives

\begin{equation}
[X/H]=(0.88\pm0.08){\:}{\rm log}{\:}\Delta v_{90} -(1.86\pm0.16),
\label{Eqn:velfitsdla}
\end{equation}

\noindent The ionization-corrected sub-DLA sample yields a very similar fit with a slope of 0.89$\pm$0.08 and an intercept of -1.90$\pm$0.16. The linear regression fit for DLAs is found to
have a slope of 1.02$\pm$0.03 and an intercept of -3.10$\pm$0.06. The slope and intercept of the fit to the combined sub-DLA and DLA sample are 1.06$\pm$0.03 and -3.00$\pm$0.06, respectively.
We note that our slope for the DLA velocity width-metallicity relation is different from that derived by \citet{Led06} and \citet{Mol13}. This difference may arise from the fact that those
studies combined metallicity measurements from depleted and undepleted elements, and did not convert all the metallicities using a uniform set of solar abundances.

The velocity width-metallicity relations for sub-DLAs and DLAs have similar slopes but their intercepts are found to be significantly different (at a $\sim 7 \sigma$ level). Similar findings
were reported by \cite{DZ09} using Fe as metallicity indicator. However, the offset between the sub-DLA and DLA trends seen in their analysis ($\sim 0.5$ dex) is smaller compared to that
found in this work ($\sim 1$ dex), reflecting the effect of depletion due to the use of Fe as metallicity indicator by \citet{DZ09}. In any case, it is clear that the offset between the two
relations exists and this may suggest different mass-metallicity relations obeyed by the galaxies traced by DLAs and sub-DLAs. This difference, together with the difference in metallicity
evolution trends, suggests that sub-DLAs and DLAs may trace distinct populations of galaxies, with sub-DLAs representing the more metal enriched galaxies.

\subsection{Comparison with Emission-line Metallicities of Absorbing Galaxies}

Nearby galaxies often show negative radial metallicity gradients. For example, Kennicutt et al. (2003) measured a gradient of -0.043 dex kpc$^{-1}$ in M101, while Rosolowky \& Simon (2008)
measured a gradient of -0.027 dex kpc$^{-1}$ in M33. The metallicity gradients show correlations with the galaxy Hubble type, bar strength and merger events (e.g., Kewley et al. 2010).
Therefore, the study of metallicity gradients in distant galaxies is of great interest for galaxy evolution models. Cresci et al. (2010) have reported `inverse' gradients in z$\sim$3 galaxies,
with the central regions having a lower metallicity than outer regions  (unlike trends in local galaxies). Such inverse gradients can result from  dilution of the  central gas by accretion of
primordial gas, as expected in cold gas accretion models (e.g., Rupke et al. 2010). DLA/sub-DLAs offer an excellent direct probe of the metallicity gradients in distant galaxies. This is
because the absorption-line metallicity pertains to the gas along the sightline to the background quasar, while the emission-line metallicity in the foreground galaxy (if known) refers to the
gas near the center of the galaxy. Thus, one can estimate the metallicity gradient in the absorbing galaxy by comparing the absorption-line metallicity and the emission-line metallicity. For
the sub-DLAs toward PHL 1226 and PKS 0439-433, measurements of emission-line metallicities of the absorbing galaxies exist in the literature. We now compare these emission-line metallicities
with the absorption-line metallicities obtained above from our HST spectra.

As mentioned in section 3.1.1, the $z=0.1602$ absorber toward PHL1226 has been identified with two spiral galaxies (G4 and G3 in the notation of Bergeron et al. 1988) at $z = 0.1592$ and
$z=0.1597$ and with impact parameters of 17.7 and 30.1 kpc, respectively. Ellison  et al. (2005) detected emission lines of H-$\beta$, [O II] $\lambda 3727$ and [O III] $\lambda 5007$ in both
galaxies. Christensen et al. (2005) obtained detections of several emission lines in galaxy G4 using integral field spectroscopy. The metallicity of galaxy G4 has been estimated by
Christensen et al. (2005) to be 12 + log (O/H) = $8.66 \pm 0.10$ from the index O3N2 $\equiv$ log ([O III] 5007/H-$\beta$) /([N II] 6583/H-$\alpha$). We note that the metallicity estimate for
G4 from the index $R_{23} \equiv$ ([O II] 3727 + [O III] 4959, 5007)/ H-$\beta$ is $8.9 \pm 0.2$ (for the upper branch of the $R_{23}$ vs. metallicity relation) from Ellison et al. (2005),
and 12 + log (O/H) = $9.02 \pm 0.13$ from Christensen et al. (2005). Given the uncertainties in the $R_{23}$ calibration, we adopt, for galaxy G4, the value 12 + log (O/H) = $8.66 \pm 0.10$
estimated from the more reliable O3N2 index (see, e.g., Pettini \& Pagel 2004). Galaxy G3, for which no O3N2 measurement is available, lies in the turnaround region of $R_{23}$; for this
galaxy, Ellison et al. (2005) estimate 12 + log (O/H) in the range $8.4^{+0.3}_{-0.1}$. This implies logarithmic emission-line metallicities relative to the solar level of [O/H] $ = -0.03 \pm
0.10$ for galaxy G4 and [O/H] $= -0.29^{+0.3}_{-0.1}$ for galaxy G3 for our assumed solar abundance 12 + log  (O/H)$_{\odot}$ =8.69 (\citealt{Asp09}). The ionization-corrected absorption-line
metallicity from our HST data [S/H] = $-0.02 \pm 0.15$ is consistent, within the uncertainties, with the emission-line metallicities of both G4 and G3. Thus either galaxy is consistent with
being the absorber. Furthermore, given the impact parameters probed by the background quasar and the consistency of the emission-based and absorption-based metallicities, there appears to be
no strong evidence of a metallicity gradient in either galaxy. Based on the difference between the absorption-line metallicity along the quasar sightline and the emission-line metallicity in
the foreground galaxy center, our estimates of the metallicity gradient are $0.0006 \pm 0.010$ dex kpc$^{-1}$ for galaxy G4 and $0.0090 \pm 0.0083$ dex kpc$^{-1}$ for galaxy G3. We note that
the uncertainties stated here include the quoted uncertainties in the absorption-line metallicity and the emission-line metallicity, combined in quadrature, but do not include uncertainties
in the ionization corrections, in the O3N2 or $R_{23}$ calibration, or in the astrometric estimation of the impact parameter. The total uncertainties in the metallicity gradients are thus
larger. We also note that the emission-line metallicities and absorption-line metallicities can trace different phases of the gas; but our data suggest the absorption-line metallicity also
traces ionized gas in this absorber. 

The $z=0.1012$ absorber toward PKS 0439-433 has been identified with a luminous ($L=0.98 L^{*}$) disk galaxy at $z=0.1010$ and with an impact parameter of 7.6 kpc. Chen et al. (2005) detected
emission lines of  H-$\alpha$, H-$\beta$, [O II] $\lambda\lambda 3726, 3729$, [O III] $\lambda \, \lambda 4959, 5007$, [N II] $\lambda \, \lambda$ 6548, 6583, and [S II] $\lambda \, \lambda$ 6716,
6731 from this galaxy. Their estimates of 12 + log (O/H) in the emitting gas were: $8.68 \pm 0.18$ based on the N2 $\equiv$ log ([N~II]/H-$\alpha$) index, $8.77 \pm 0.14$ based on the O3N2
index, and $9.19 \pm 0.15$ based on the $R_{23}$ index.  For our assumed solar abundances, these estimates imply emission-line metallicities of [O/H] = $-0.01 \pm 0.18$, $0.08 \pm 0.14$, and
$0.42 \pm 0.15$, respectively. All of these estimates of the emission-line metallicity are consistent with the absorption-line metallicity derived from our HST data, i.e., [S/H] = $0.10 \pm
0.15$ after ionization correction; this suggests that the metallicity gradient is not significant in this galaxy. Based on the difference between the absorption-line metallicity along the
quasar sightline and the O3N2 estimate of the emission-line metallicity in the foreground galaxy center, our estimate of the metallicity gradient in this galaxy is $0.0026 \pm 0.027$ dex
kpc$^{-1}$. We note that Chen et al. (2005) had suggested the existence of a metallicity gradient in this galaxy  based on their estimate of [Fe/H] from low-resolution HST FOS spectra. We
believe that our assessment of the absence of a significant abundance gradient in this galaxy is likely to be more correct, because of the higher resolution of our HST COS spectra compared to
the FOS spectra, our use of the nearly undepleted $\alpha$-element S rather than the heavily depleted element Fe, and the likelihood of the emission-line metallicity to be closer to the O3N2
value rather than the $R_{23}$ value. We note again that the emission-line metallicities and absorption-line metallicities can trace different phases of the gas; but our data suggest the
absorption-line metallicity also traces ionized gas in this absorber.

More accurate determinations of the emission-line metallicities and absorption-line metallicities are essential to examine whether these galaxies could possess weak metallicity gradients
(e.g., -0.022 dex kpc$^{-1}$ suggested by Christensen et al. 2014). We also note that uncertainties in the ionization correction, emission-line calibration, and galaxy center determinations
should also be taken into account. But we point out that the lack of a significant metallicity gradient is consistent with the suggestion of P\'eroux et al. (2014) that abundances derived in
absorption along the quasar sightline are reliable indicators of the overall galaxy metallicities and, furthermore, that comparison of emission-line and absorption-line  metallicities does
not support the infall of fresh, metal-poor gas in these objects.

\section{CONCLUSIONS}

Our observations have doubled the sample of sub-DLAs at $z < 0.5$ with metallicity measurements. Moreover, they have increased the sample of S detections (rather than upper limits) for
sub-DLAs at $z < 0.5$ by a factor of 4. Our main conclusions are as follows:

1. The available data are consistent with all four of our sub-DLAs having near-solar or super-solar metallicity. Observations of more absorption lines and detailed modeling will help to
further verify this.

2. Simple photoionization calculations suggest that while there is a significant amount of ionized gas in some of our absorbers, the ionization corrections to the element abundances are
relatively modest (typically $\la 0.2$ dex).

3. Our data suggest that the H~I column density-weighted mean metallicity of sub-DLAs is higher than that of DLAs at $z\sim0$. Furthermore, in combination with the measurements from the
literature, our data confirm previous findings that sub-DLAs, on average are more metal-rich than DLAs at all redshifts studied. Moreover, we have shown that the observed difference between
sub-DLA and DLA metallicities is not likely due to ionization effects.

4. There appears to be significant dust depletion in the absorber toward PHL 1598, in which the abundances of both C and Si are considerably sub-solar compared to S. 

5. In the absorber toward PKS 0439-433, the ratio Mn/Fe appears to be super-solar, far above the [Mn/Fe] vs. metallicity trend observed for higher-redshift sub-DLAs. There is no strong
evidence for an odd-even effect between S and P. 

6. The abundance of N is considerably lower than that of S, and in fact even below the level expected for secondary N production for several sub-DLAs at $z < 0.5$. This may suggest a delay
in the release of the secondary N or a tertiary N nucleosynthesis mechanism.

7. We report possible detections of Si II$^{*}$ in two sub-DLAs. These, if real, would be the first Si II$^{*}$ detections in intervening quasar sub-DLAs, to our knowledge. Using these, we
constrain the electron densities in the absorbers toward PHL 1226 and PKS 0439-433 to be in the range $\approx 0.8-1.3$ cm$^{-3}$ and $< 8.0-13.9$ cm$^{-3}$, respectively. 

8. Based on limits to the C II$^{*}$/ C II ratio, the absorber toward TXS 0454-220 does not appear to be significantly ionized. By contrast, the absorbers toward PHL 1226 and PKS 0439-433
show significant amounts of ionized gas, larger velocity widths, and asymmetric O VI profiles, suggesting that outflows may play an important role in these objects. 

9. The velocity dispersion ($\Delta v_{90}$) vs. metallicity data for DLAs and sub-DLAs taken together appear to show a correlation suggesting that more massive absorber galaxies are likely
to be more metal-rich as well. Furthermore, the $\Delta v_{90}$ vs. metallicity relations for sub-DLAs and DLAs appear to be different from each other. If $\Delta v_{90}$ is an indicator of
the mass of the absorbing galaxy, then our finding suggests that the populations of galaxies traced by DLAs and sub-DLAs obey different mass-metallicity relations and that sub-DLAs are likely
to trace more massive galaxies than DLAs.

10. For the sub-DLAs toward PHL 1226 and PKS 0439-433, our measurements of the absorption-line metallicities along the quasar sightlines are consistent with measurements, from the literature,
of the emission-line metallicities near the centers of the foreground absorbing galaxies. This suggests that metallicity gradients are not significant in these galaxies. 

Future higher resolution UV spectroscopy covering weaker lines of key species such as O I as well as higher ions such as O VI, S IV, S VI would prove extremely useful in further constraining
element abundances and ionization in multi-phase gas in our low-redshift sub-DLAs. Higher S/N spectra would also allow more accurate determinations of the H I column densities; this is
especially important since the uncertainties in H I column densities, if large, can dominate the uncertainties in the element abundances derived. Moreover, observations of higher Lyman series
lines would be extremely valuable for constraining the H~I column densities in individual velocity components, which would allow more sophisticated photoionization modeling. Finally, we note
that the samples of sub-DLAs and DLAs at $z < 0.6$ are still much smaller than the samples at $z > 0.6$. Observations of more sub-DLAs and DLAs at low redshifts are essential to shed further
light on how these systems compare with their high-$z$ analogs and with the Milky Way ISM, and how they evolve with time.

\acknowledgments
This work was supported in part by a grant from NASA/Space Telescope Science Institute for program GO 12536 (PI Kulkarni). Additional support from the US National Science Foundation grant
AST/1108830 and NASA grant NNX14AG74G (PI Kulkarni) is gratefully acknowledged. We thank B. DeMarcy for assistance with some of the photoionization simulations and T. Zafar for making
available her compilation of N abundance measurements ahead of publication. We are thankful to an anonymous referee for comments that have helped to improve this paper.

{\it Facilities:} \facility{HST (COS)}.

\clearpage

\begin{figure}
\includegraphics[angle=90,scale=.30]{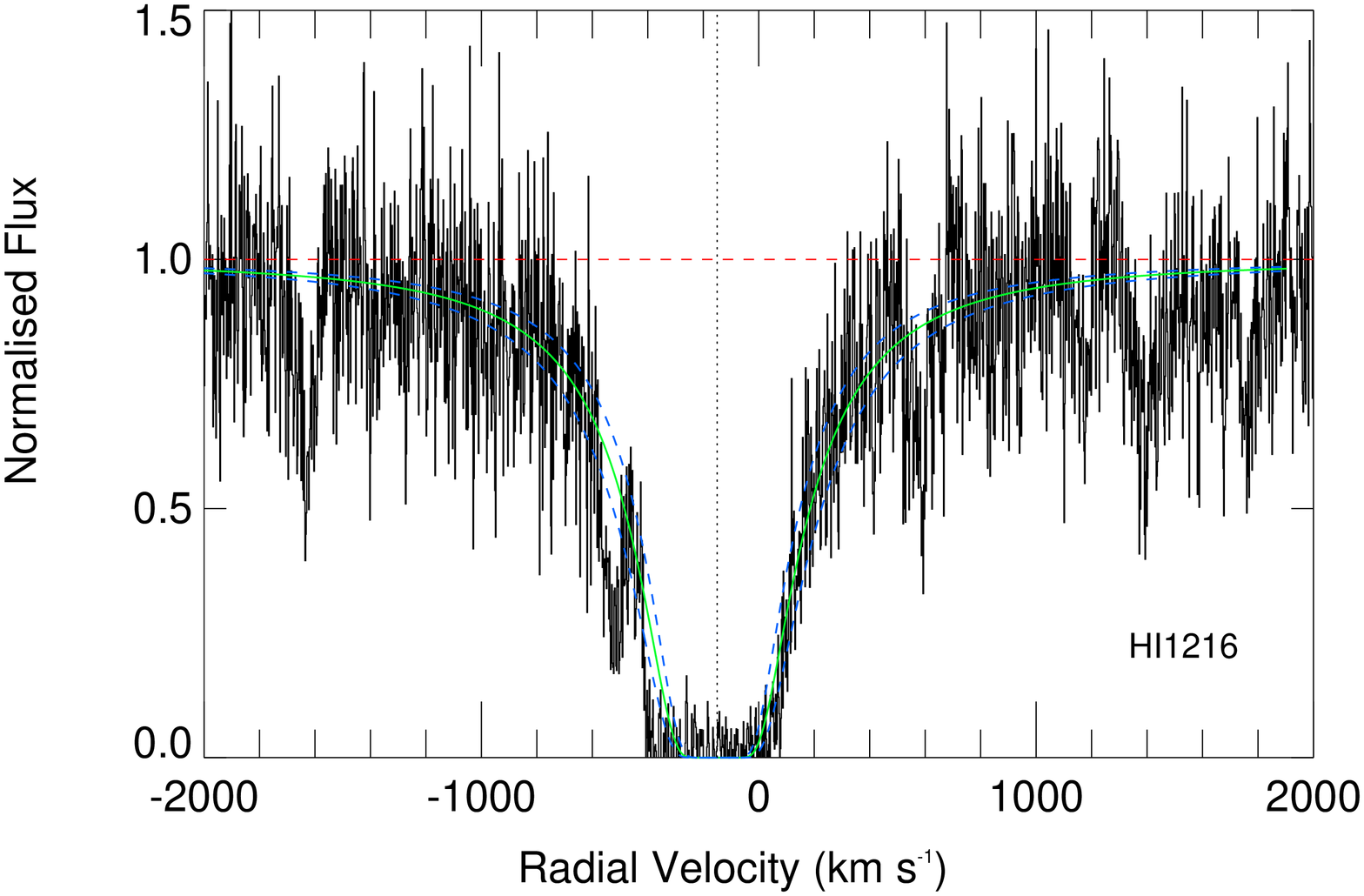}
\includegraphics[angle=90,scale=.30]{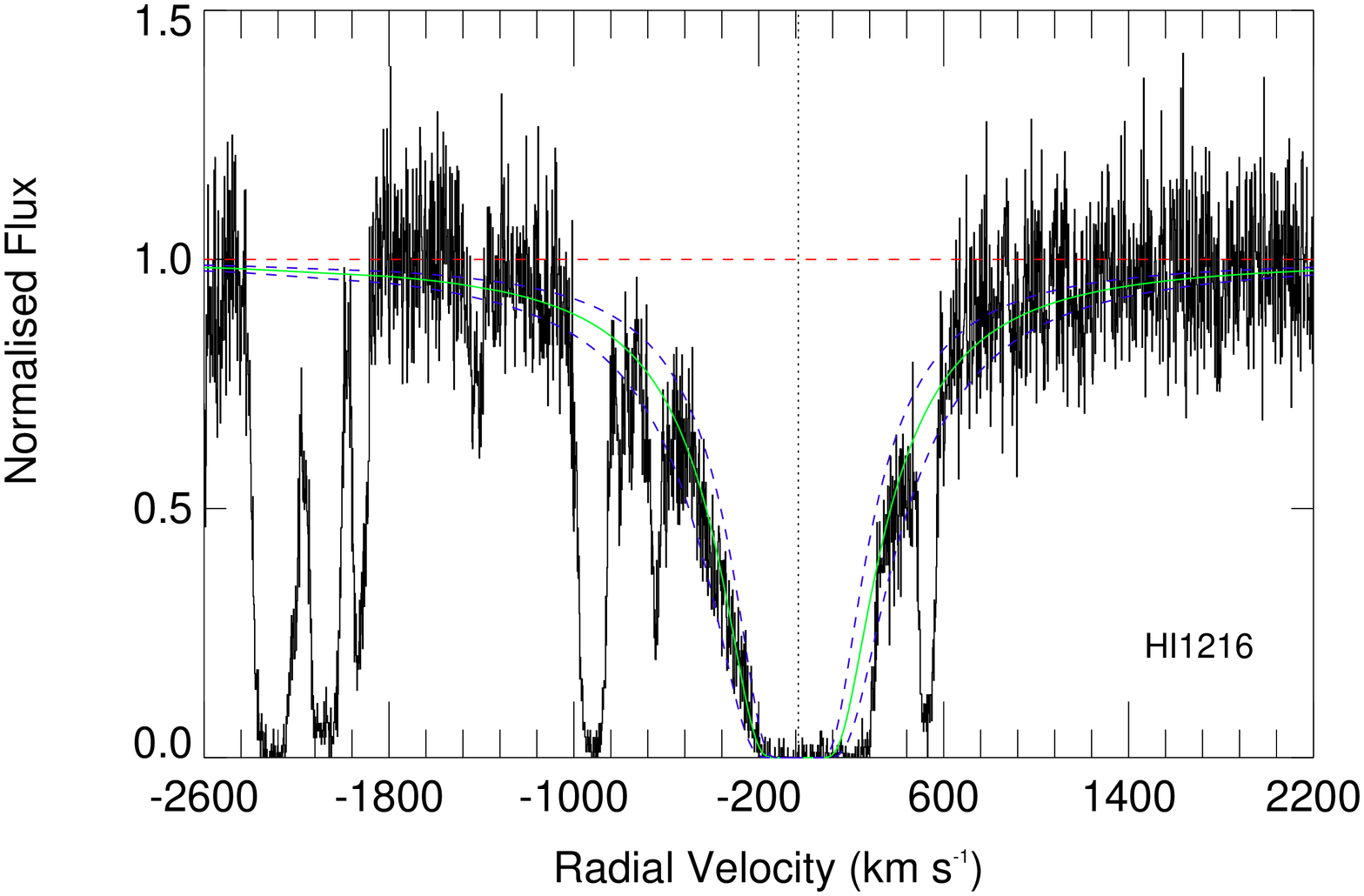}
\caption{H I Ly-$\alpha$ absorption feature in our HST COS spectra for (a) the $z=0.1602$ absorber toward PHL 1226 (left), and (b) the $z=0.1012$ absorber toward PKS 0439-433 (right). In each panel, the solid green curve denotes the best-fitting Voigt profile and the dashed blue curves denote the profiles corresponding to estimated $\pm 1\sigma$ deviations from the best-fitting $N_{\rm H I}$ value. The horizontal dashed red line denotes the continuum level. The vertical dotted black line denotes the center of the Ly-$\alpha$ line.  
\label{fig1}}
\end{figure}

\begin{figure}
\includegraphics[angle=90,scale=.60]{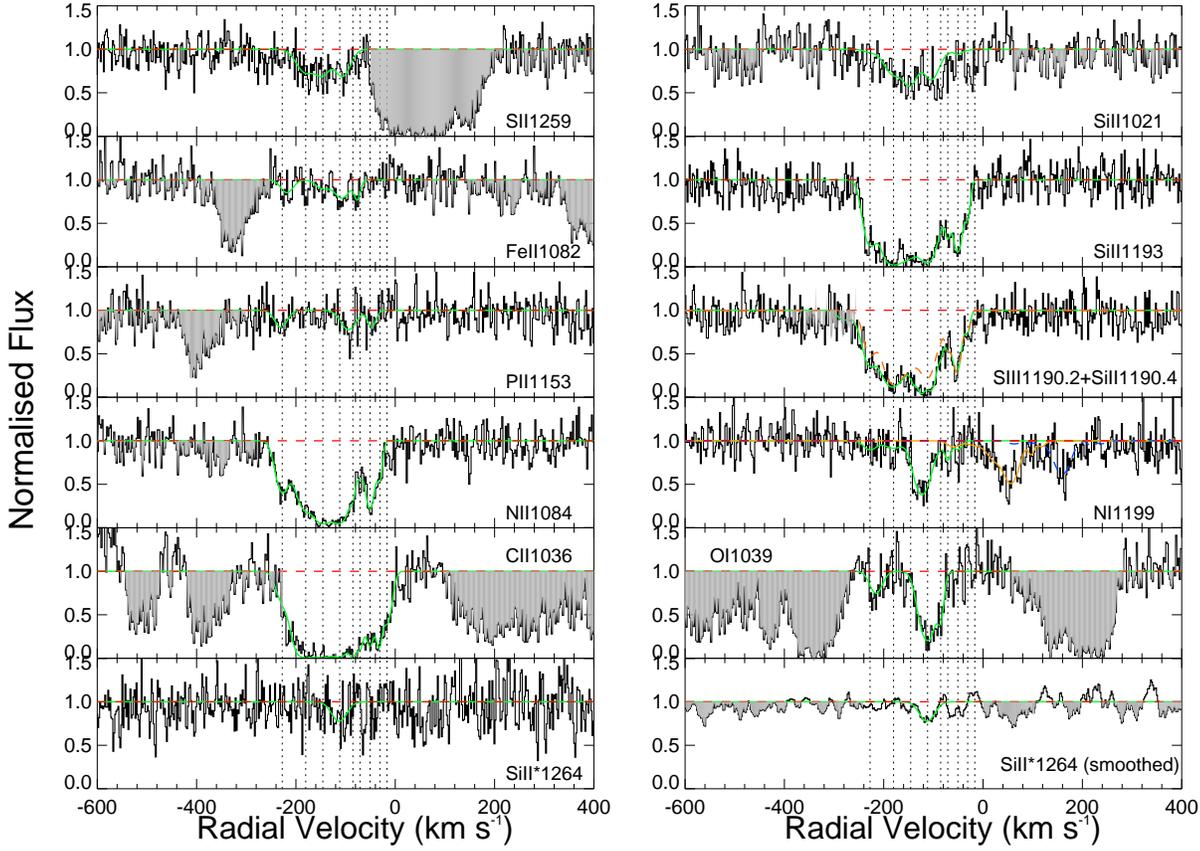}
\caption{ Velocity plots for several lines of interest for the  $z =0.1602$ system in the spectrum of PHL 1226. The solid green curve indicates the theoretical Voigt profile fit to the
absorption feature, and the dashed red line is the continuum level. The vertical dotted lines indicate the positions of the components that were used in the fit. In the S III 1190.2 + Si II
1190.4 panel (with velocity scale shown for  Si II $\lambda 1190.4$), the solid green curve represents the combined contributions from S III $\lambda$ 1190.2 and Si II $\lambda$ 1190.4 lines
while the contribution from Si~II $\lambda$ 1190.4 alone to this blend, as determined from the Si II $\lambda$ 1193.3 line, is represented by the dashed orange curve. We obtained an upper
limit on the column density of S III by fitting the rest of the blend. The simultaneous fits to the N I $\lambda\lambda$ 1199.6, 1200.2, 1200.7 lines are shown in the N I 1199 panel using
solid green, solid orange and dashed blue curves, respectively, and the velocity scale is shown for N I $\lambda$ 1199.6. The bottom panels in both columns show possible absorption from the
Si II$^{*}$ $\lambda 1264.7$ line, with the panel on the right presenting the data smoothed by a factor of 9. The regions shaded in gray in some of the panels represent absorption unrelated
to the line presented. As noted in section 2.3, the Voigt profile fits provide an approximate description of the velocity structures of the absorbing gas, but yield accurate total column
density estimates.
\label{fig2}}
\end{figure}

\begin{figure}
\includegraphics[angle=90,scale=.60]{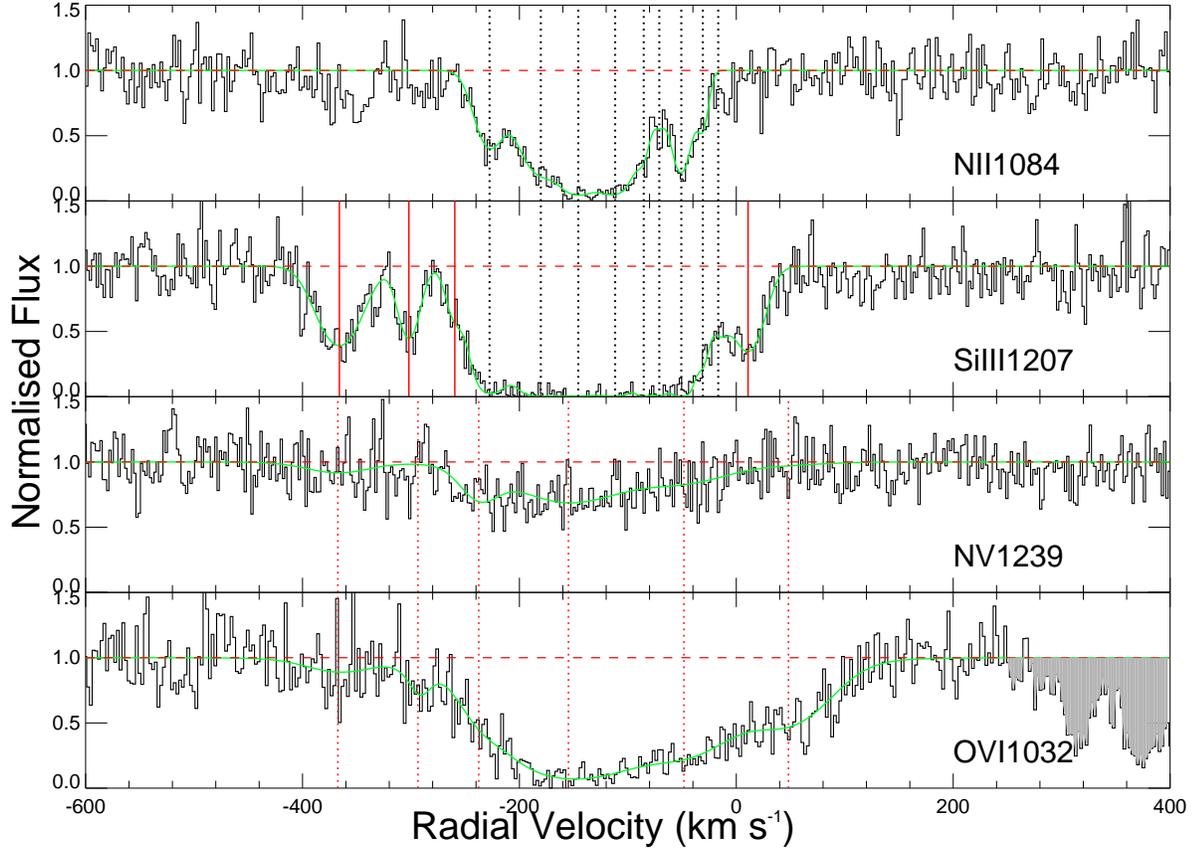}
\caption{Same as Fig. 2, but for Si III, N V and O VI absorption in the $z =0.1602$ system in the spectrum of PHL 1226. The N II $\lambda$ 1084 absorption profile is replotted in the
top panel for comparison. Positions of the components used to model the absorption from Si~III, N II, as well as other low-ions (see Fig. \ref{fig2}) are indicated by vertical black dotted
lines. The solid vertical red lines in the Si~III panel mark the positions of additional absorption components seen in Si III $\lambda$ 1207. The fits to the N V $\lambda$ 1239 and O VI
$\lambda$ 1032 lines adopt a velocity structure which is independent of that used for the lower-ion transitions. The dotted vertical red lines indicate the positions of the components used to
model N V and O VI absorption. 
\label{fig3}}
\end{figure}

\begin{figure}
\includegraphics[angle=90,scale=.60]{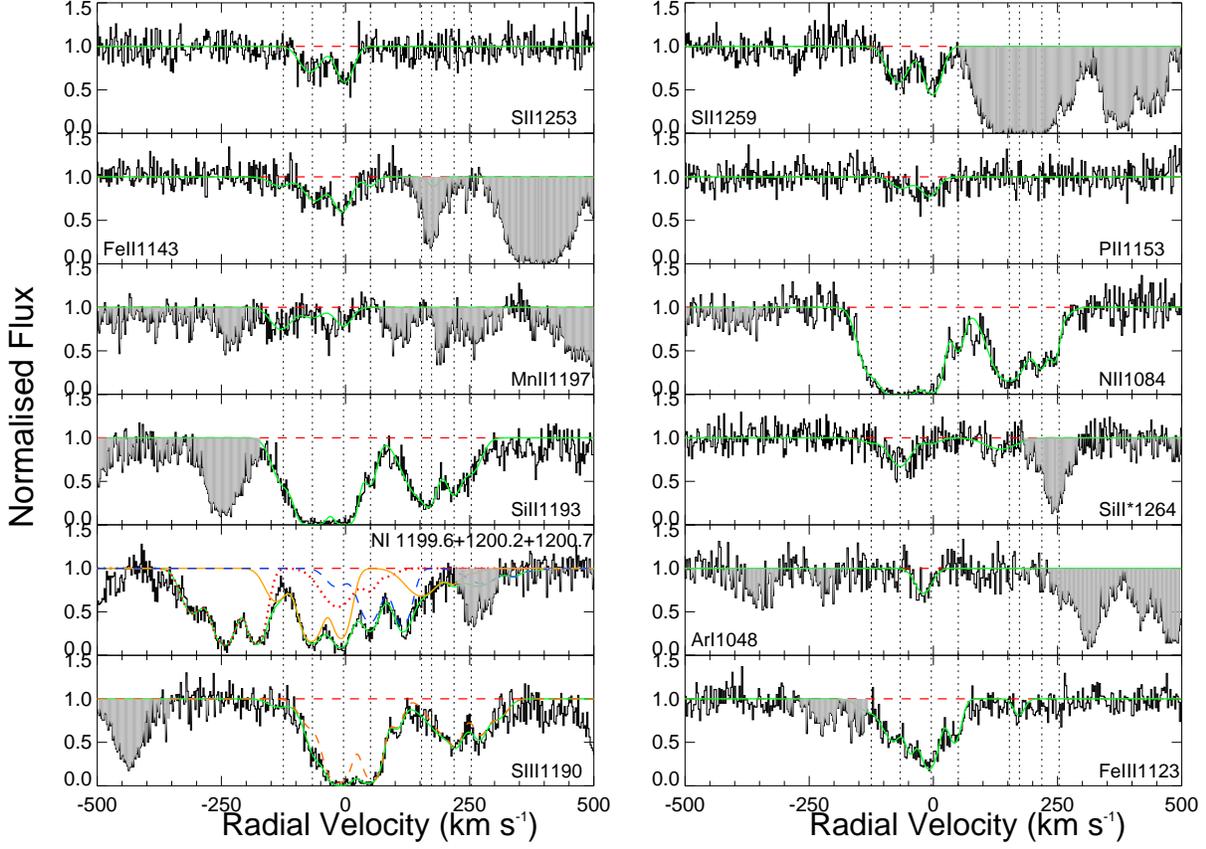}
\caption{Same as Fig. 2, but for the $z =0.1012$ sub-DLA absorption system in the spectrum of PKS 0439-433.  In the N I 1199.6+1200.2+1200.7 panel, the solid green curve represents the total contribution from the N I $\lambda\lambda$ 1199.6, 1200.2, 1200.7 lines. The individual fits to N~I $\lambda$ 1199.6, N I $\lambda$ 1200.2 and N I $\lambda$ 1200.7 are shown using the dotted red, solid orange, and dashed blue curves, respectively, and the velocity scale is shown for N~I $\lambda$ 1200.2. The solid green curve in the S III 1190 panel (with velocity scale shown for  S III $\lambda 1190.2$) represents the combined contributions from S III $\lambda$ 1190.2 and Si II $\lambda$ 1190.4 lines while the contribution from Si II $\lambda$ 1190.4 alone to this blend, as determined from the Si II $\lambda$ 1193.3 line, is represented by the dashed orange curve. We obtained an upper limit on the column density of S III by fitting the rest of the blend. 
\label{fig4}}
\end{figure}

\begin{figure}
\includegraphics[angle=90,scale=.60]{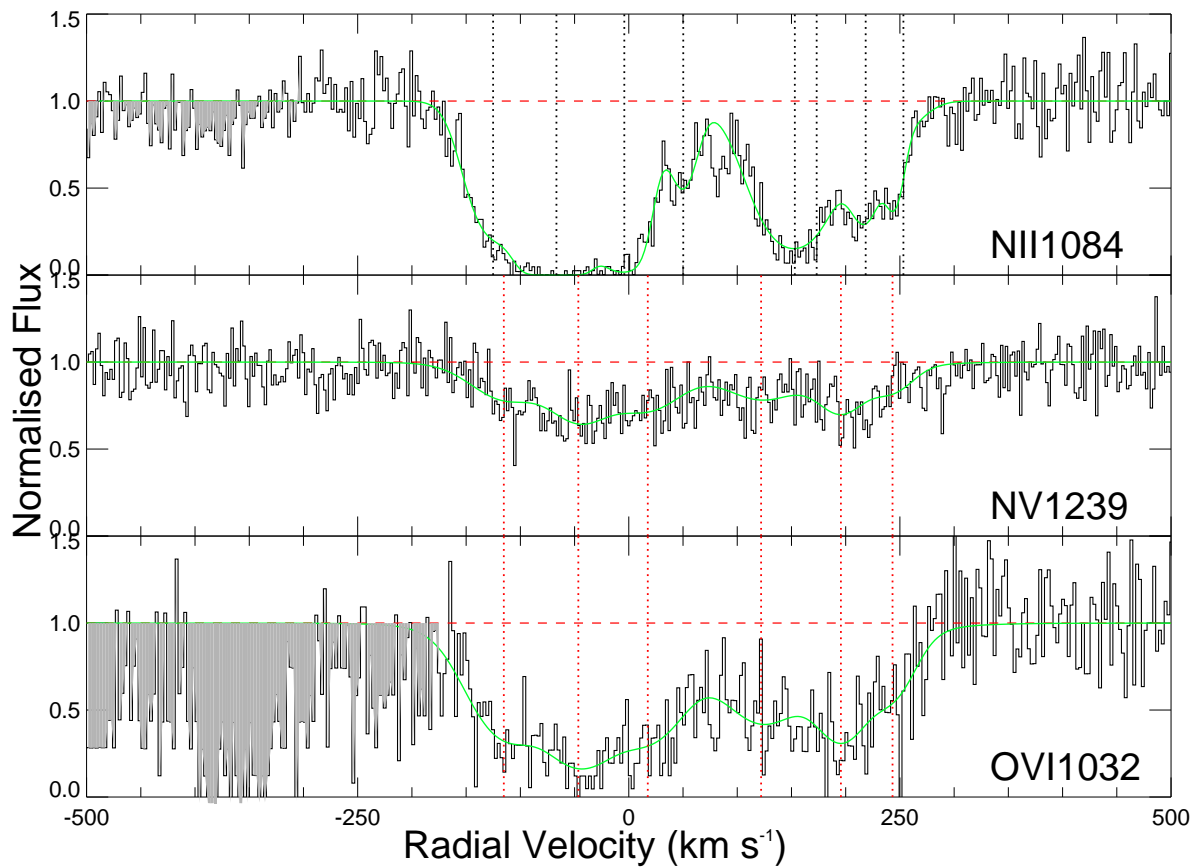}
\caption{Same as Fig. \ref{fig2}, but for N V and O VI absorption in the $z =0.1012$ system in the spectrum of PKS 0439-433. The N II $\lambda$ 1084 absorption profile is replotted in the
top panel for comparison. The vertical dotted black lines represent the locations of the components used in the fit to the N~II line as well as other low-ion transitions. The positions of the
components used to model the N V $\lambda$ 1239 and O VI $\lambda$ 1032 lines are marked by vertical red dotted lines.
\label{fig5}}
\end{figure}

\begin{figure}
\includegraphics[angle=90,scale=.60]{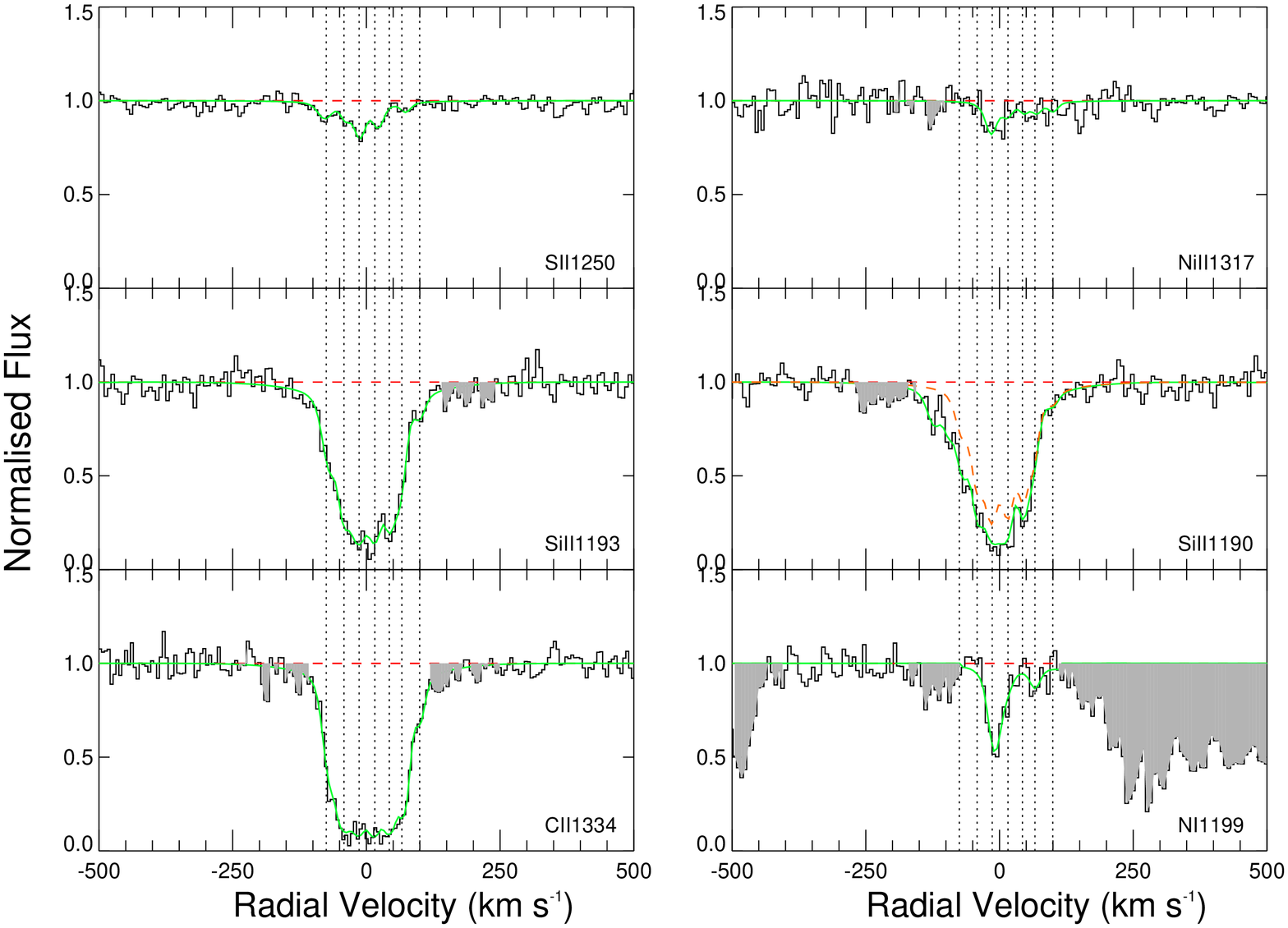}
\caption{Same as Fig. 2, but for the z =0.4744 system in the spectrum of TXS 0454-220. The solid green curve in the Si II 1190 panel (with velocity scale shown for  Si II $\lambda 1190.4$) represents the combined contributions from S III $\lambda$ 1190.2 and Si II $\lambda$ 1190.4, while the contribution from Si II $\lambda$ 1190.4 alone to this blend, as determined from Si II $\lambda$ 1193.3, is represented by the dashed orange curve. We obtained an upper limit on the column density of S III by fitting the rest of the blend. \label{fig6}}
\end{figure}

\begin{figure}
\includegraphics[angle=90,scale=.60]{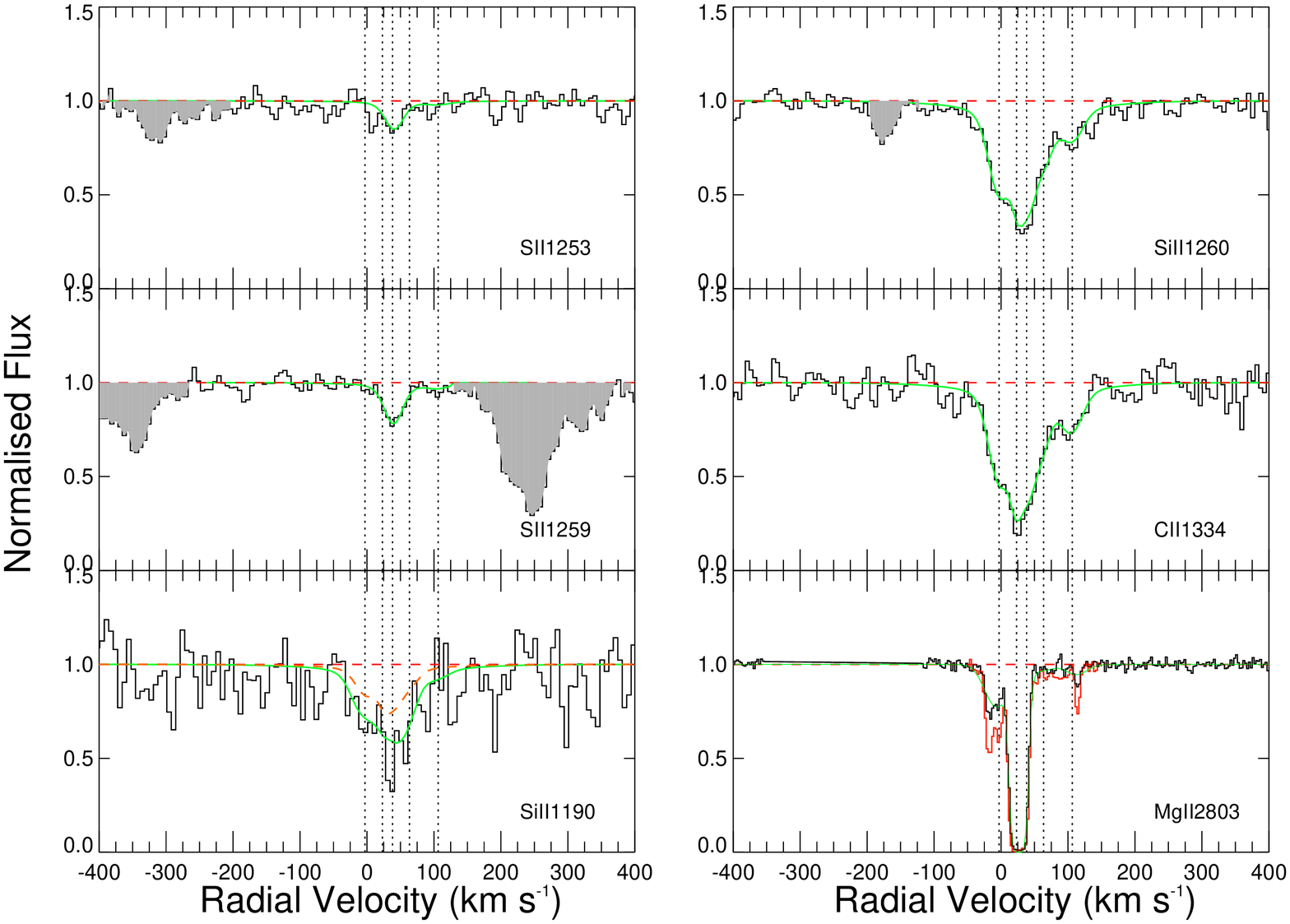}
\caption{ Same as Fig. 2, but  for the  $z =0.4297$ system in the spectrum of PHL 1598. The solid green curve in the Si II 1190 panel (with velocity scale shown for Si II $\lambda 1190.4$)
represents the combined contributions from S III $\lambda$ 1190.2 and Si II $\lambda$ 1190.4, while the contribution from Si II $\lambda$ 1190.4 alone to this blend, as determined from Si II
$\lambda$ 1260.4, is represented by the dashed orange curve. We obtained an upper limit on the column density of S III by fitting the rest of the blend. The bottom right panel shows the Mg II
$\lambda$ 2803 line from archival Keck HIRES (Program ID: C99H, PI: C. Steidel) data, resampled to match the spectral resolution of our COS-NUV data. The corresponding Voigt profile fit,
shown using the solid green line, uses the same velocity structure used to fit the lines detected in our COS-NUV spectra for this sightline. The Mg II $\lambda$ 2796 line from the resampled
HIRES spectra is also shown using the solid red line. \label{fig7}}
\end{figure}

\begin{figure}
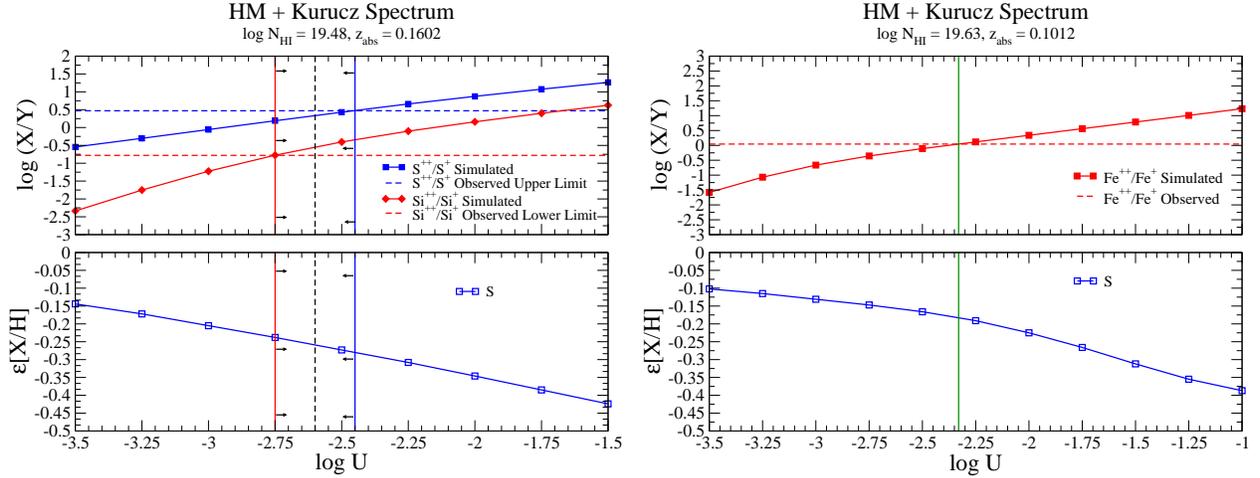

\includegraphics[angle=0,scale=.33]{PHL1226_ioniz.eps}
\hskip 0.05in
\includegraphics[angle=0,scale=.33]{Q0439_ioniz.eps}
\caption{ Results of photoionization simulations for the sub-DLAs toward (a) (Left) PHL 1226 and (b) (Right) PKS 0439-433. For each system, the top panel shows the variation with ionization parameter of the column density ratio (S III/S II and Si III/Si II for PHL 1226, and Fe III/ Fe II for PKS 0439-433). The horizontal dashed lines show the  the observed column density ratio values or limits. The vertical solid lines show the corresponding values or limits on the ionization parameter. In cases where the  ionization parameter is bounded on both sides, the dashed vertical lines represent the mean of these bounds. The bottom panels show the ionization correction for S as a function of the ionization parameter. As in the top panels, the values or bounds on the ionization parameter implied by the observed column density ratios are denoted by vertical lines. 
\label{fig8}}
\end{figure}

\begin{figure}
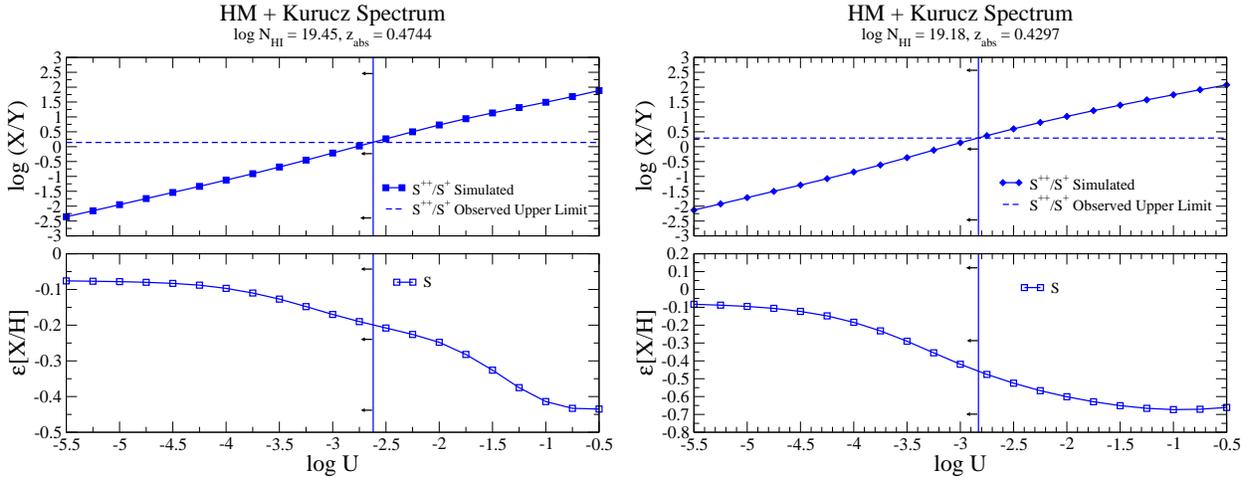

\includegraphics[angle=0,scale=.33]{Q0454_ioniz.eps}
\hskip 0.05in
\includegraphics[angle=0,scale=.33]{PHL1598_ioniz.eps}
\caption{ Results of photoionization simulations for the sub-DLAs toward (a) (Left) TXS 0454-220 and (b) (Right) PHL 1598. For each system, the top panel shows the variation with ionization parameter of the column density ratio S III/S II. The horizontal dashed lines show the observed column density ratio limits. The vertical solid lines show the corresponding upper limits on the ionization parameter. 
The bottom panels show the ionization correction for S as a function of the ionization parameter. As in the top panels, the limits on the ionization parameter implied by the observed column density ratios are denoted by vertical lines. 
\label{fig9}}
\end{figure}

\begin{figure}
\includegraphics[angle=0,scale=0.70]{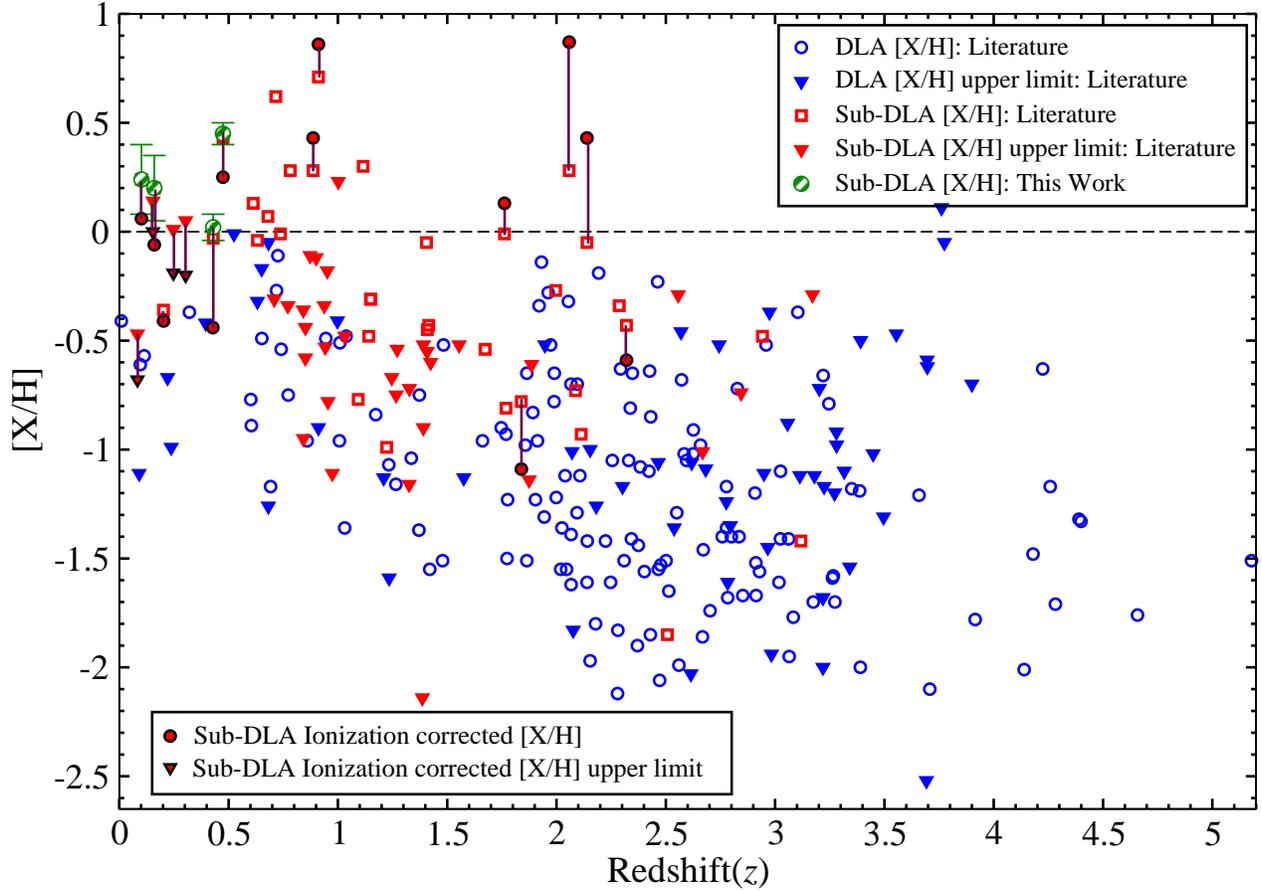}
\vskip 0.0in
\caption{Zn or S based metallicity vs. redshift data for sub-DLAs and DLAs. The sub-DLA and DLA samples contain 72 and 195 systems, respectively. The sub-DLA metallicity sample at $z \la 0.5$
contain nine [S/H] measurements, out of which five are based on the detection of S II lines (including the measurements presented in this work; represented by green striped circles) and the
rest are [S/H] upper limits. Also shown are the ionization corrected metallicity values for the sub-DLAs with available ionization correction estimates (in the literature as well as from this
work). For these sub-DLAs, the observed metallicity is connected with the corresponding ionization-corrected value by a brown vertical line to indicate the amount of correction applied.
\label{fig:scatter}}
\end{figure}

\begin{figure}
\includegraphics[angle=0,scale=0.62]{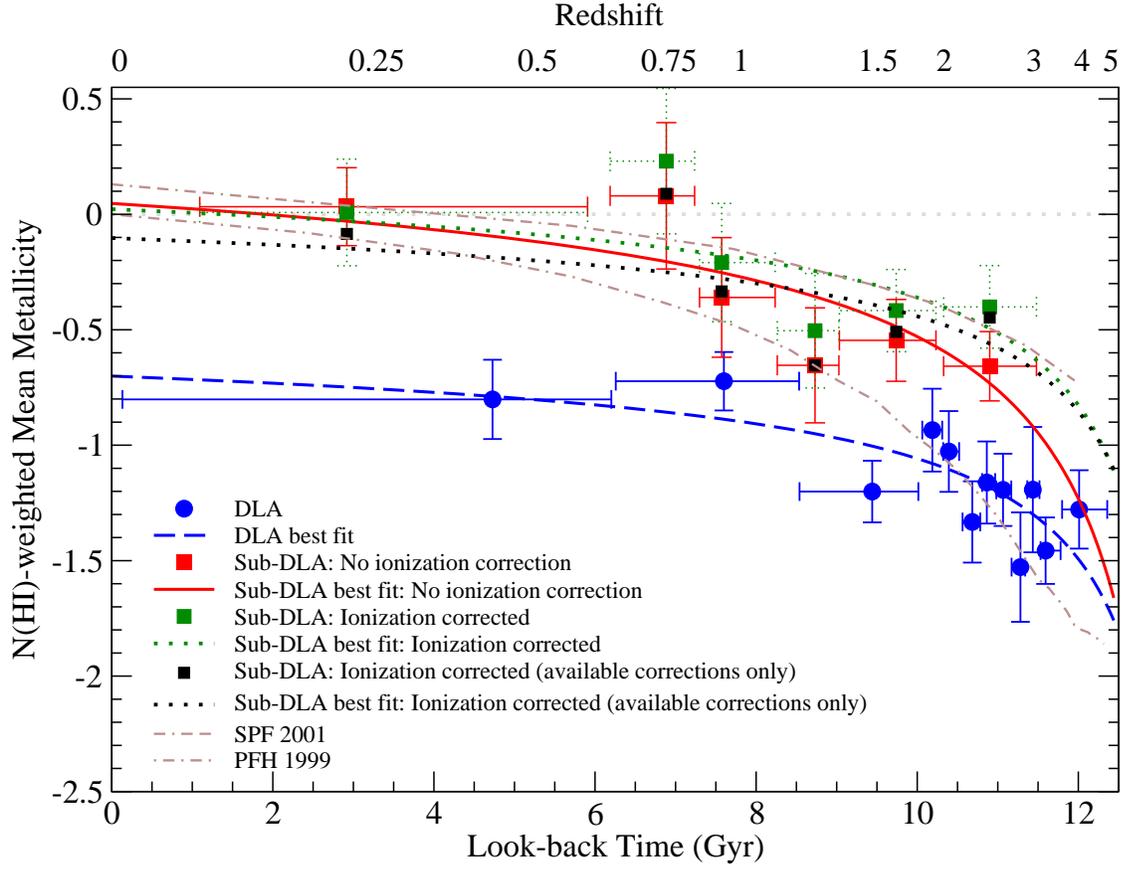}
\vskip -0.2in
\caption{H I column density-weighted mean metallicity vs. look-back time for the sub-DLA and DLAs samples shown in Fig. \ref{fig:scatter}. The data are binned, with 11-13 systems per sub-DLA
bin and 16-17 systems per DLA bin. The horizontal dotted grey line denotes the solar level. The solid red and dashed blue curves show best-fitting linear regression fits to the $N_{\rm
H I}$-weighted mean observed metallicity vs. median redshift relation for sub-DLAs and DLAs, respectively. The dot-dashed brown curve shows the prediction for the mean interstellar
metallicity in the Pei et al. (1999) model with the optimum fit for the cosmic infrared background intensity. The dot-double-dashed brown curve shows the mean metallicity of cold interstellar
gas in the semi-analytic model of  Somerville et al. (2001). Also shown are the $N_{\rm H I}$-weighted mean ionization-corrected metallicity vs. median redshift relations for sub-DLAs
considering only the available corrections (bins represented using black filled squares) and overall (available corrections + median of available S and Zn corrections applied to systems
for which corrections are unknown; bins shown using green filled squares) correction. The sub-DLAs appear to be more metal-rich than DLAs at all redshifts studied, and appear to agree better
with the models shown.
\label{fig:bin}}
\end{figure}

\begin{figure}
\includegraphics[angle=0,scale=.7]{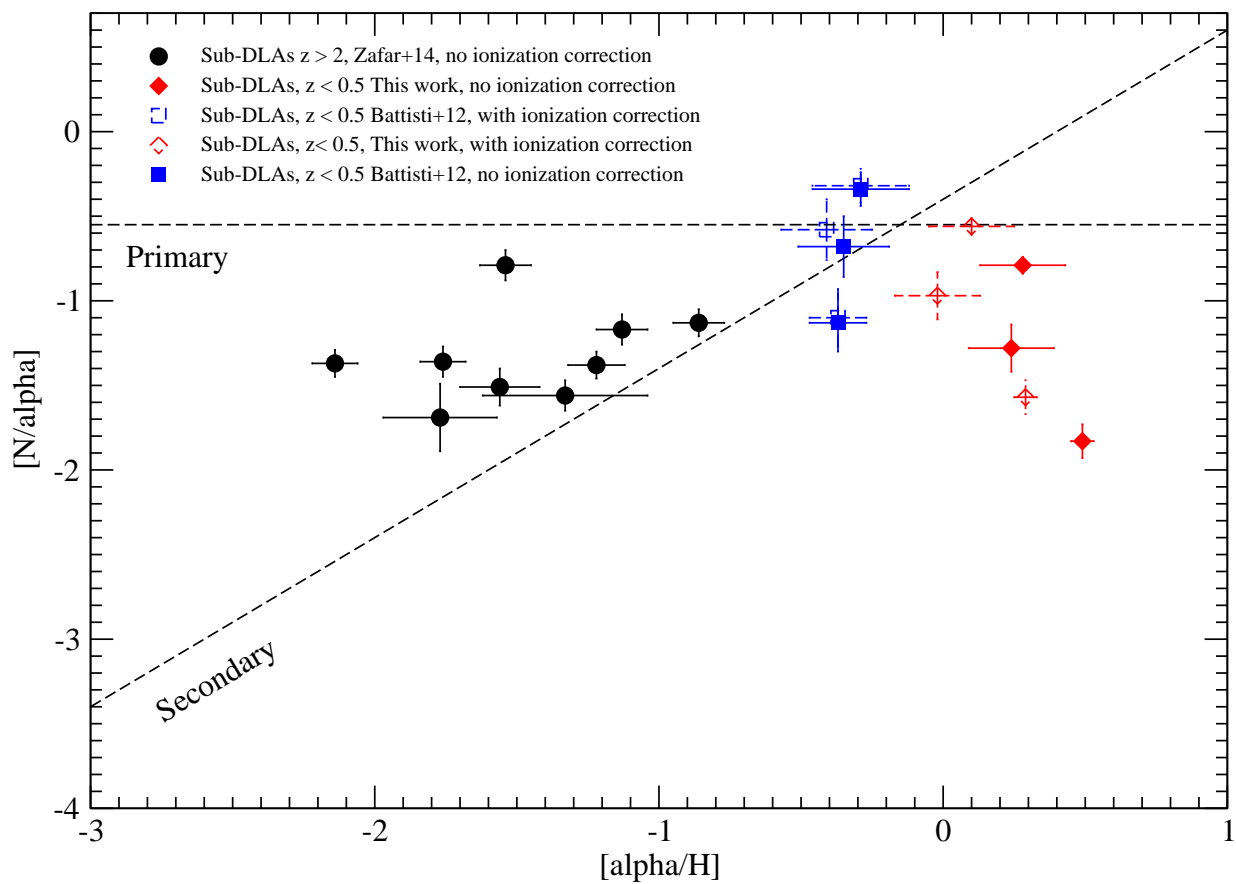}
\caption{ Observed abundance of N relative to an $\alpha$ element (S or O) vs. the observed abundance of that $\alpha$ element in 
sub-DLA absorbers at $z > 2$ and at $z < 0.5$, based on measurements presented here, Battisti et al. (2012), and 
Zafar et al. (2014). The dashed lines show approximate levels for primary and secondary N production. 
\label{fig13}}
\end{figure}

\begin{figure}
\includegraphics[angle=0,width=\linewidth]{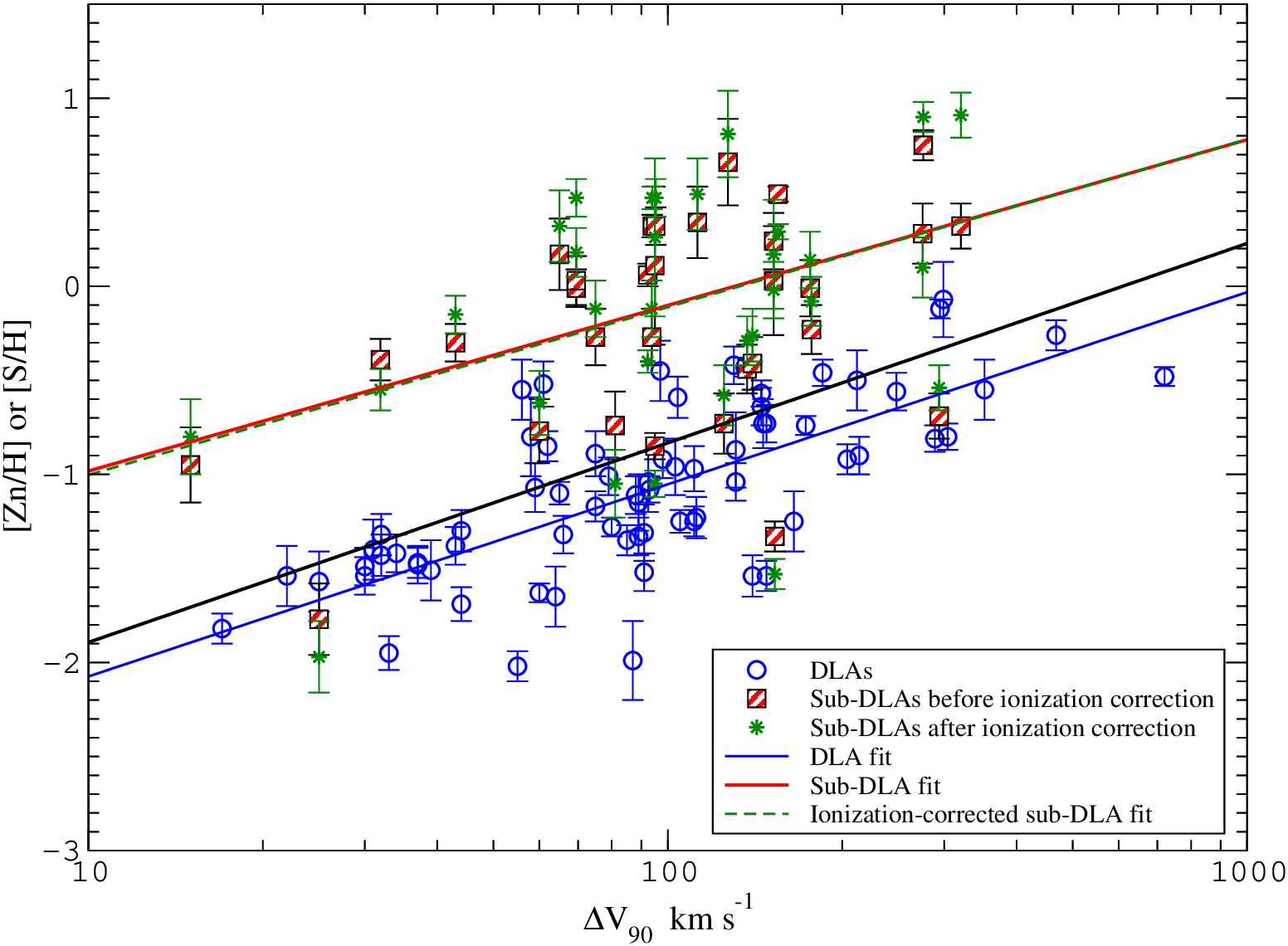}
\vskip 0.0in
\caption[$\Delta$V vs. {[Zn/H]} relation]{Velocity width ($\Delta V_{90}$) vs. Metallicity relations for sub-DLAs and DLAs. The ionization-corrected sub-DLA data are shown as well. Linear
regression fits to the sub-DLA and DLA data are shown with solid red and blue lines respectively. The fit for the ionization-corrected sub-DLA data is shown with a dashed green line. The
solid black line represents the linear regression fit to the combined (DLA + sub-DLA) sample. \label{Fig:kinematicsfig}}
\end{figure}

\clearpage

\begin{table}
\begin{center}
\caption{Targets Observed}
\vskip 5pt
\begin{tabular}{lcccccc}
\tableline\tableline
Quasar& $z_{quasar}$&$z_{abs}$&log $N_{\rm H I}^{1}$&$z_{gal}^{2}$&$D^{3}$&Reference\\
&&&&&&(abs, em)$^{5}$\\
\tableline
PHL 1226&0.404&0.1602&$19.48 \pm 0.10$&$0.1592, 0.1597$&17.7, 30.1 & 1, 3 \\
PKS 0439-433&0.593&0.1012&$19.63\pm0.15$&0.1010&7.8&1, 4\\
TXS 0454-220&0.534&0.4744&$19.45\pm0.03$&...&...&2,...\\
PHL 1598&0.501&0.4297&$19.18\pm0.03$&0.4299&48.2&2, 5\\
\tableline
\end{tabular}
\tablenotetext{1}{HI column densities reported in the literature based on prior lower-resolution HST data or based on our 
HST data, in cm$^{-2}$.}
\tablenotetext{2}{Spectroscopic emission redshift of candidate absorber galaxy, if known.}
\tablenotetext{3}{ Impact parameter in kpc of the quasar sightline from the galaxy center.}
\tablenotetext{4}{ References for absorber $N_{\rm H I}$: (1) This study; (2) Rao et al. (2006). 
References for galaxy identification: (3) Bergeron et al. (1988); (4) Petitjean et al. (1996); (5) Bergeron (1986)}
\end{center}
\end{table}

\begin{table}
\begin{center}
\caption{Summary of HST COS Observations}
\vskip 5pt
\begin{tabular}{cccc}
\tableline\tableline
Quasar& Grating&$\lambda_{central}$(\AA)&$t_{exp}(s) \times n_{exp}$\\
\tableline
PHL 1226&G130M&1327&$(810\times4)+ (1412 \times 3) + (1413 \times 4)$\\
PKS 0439-433&G130M&1291&$(948\times2) + (1473\times8)$\\
TXS 0454-220&G185M&1850&$(1076\times4)+(1464\times14)$\\
PHL 1598&G185M&1786&$(1089\times4)+(1462\times13)$\\
\tableline
\end{tabular}
\end{center}
\end{table}

\begin{landscape}
\begin{table}
\begin{center}
\caption{Rest-frame Equivalent Width Measurements in m{\AA} \label{ewtable}}
\vskip 5pt
\begin{tabular}{lcccccccccc}
\tableline\tableline

Quasar &$z_{abs}$& C II &C II&N I  & N II& N V & N V& O I  & O VI & Ar I\\
&&1036& 1334& 1199&1084&1239&1243&1039&1032&1048\\
\tableline
PHL 1226&0.1602&$677 \pm 36$&...&...&$569\pm14$&$301\pm98$&$174\pm87$&$171\pm57$&$929\pm128$&...\\
PKS 0439-433&0.1012&...&...&...&$1094 \pm 23$&$388\pm55$&...&...&$888\pm52$&$33\pm10$\\
TXS 0454-220&0.4744&...&$727\pm13$&$78\pm8$&...&...&...&...&...&...\\
PHL 1598 & 0.4297&...&$310\pm13$&...&...&...&...&...&...&...\\
\tableline
\end{tabular}
\vskip 10pt

\begin{tabular}{lcccccccccc}
\tableline\tableline
Quasar&$z_{abs}$& Si II & Si II  & Si II  & Si II* & Si III & P II &S II  & S II  & S II \\
&&1021&1193&1260&1265&1207 & 1153&1250&1253&1259\\
\tableline
PHL 1226&0.1602&$162\pm24$&$690\pm12$&$838\pm15$&$30\pm11$&$1192 \pm 28$&$97\pm38$&...&...&$122\pm19$\\
PKS 0439-433&0.1012&...&$1039\pm30$&$1279\pm22$&$136 \pm 42$&...&$67\pm22$&$77\pm18$&$131\pm16$&$177\pm15$\\
TXS 0454-220&0.4744&...&$534\pm14$&$747\pm7$&...&...&...&$68\pm8$&$...$&$111\pm4$\\
PHL 1598 & 0.4297&...&...&$266 \pm 6$&...&...&...&...&$32\pm7$&$36\pm6$\\

\tableline
\end{tabular}
\vskip 10pt

\begin{tabular}{lcccccccccc}
\tableline\tableline

Quasar &$z_{abs}$&Mn II & Fe II  & Fe II & Fe II & Fe II& Fe II& Fe II&Fe III& Ni II \\
&& 1197&1082&1097&1112&1125& 1143& 1145&1123& 1317\\
\tableline
PHL 1226&0.1602&$<10$&$65\pm23$&...&$137\pm13$&...&...&...&...\\
PKS 0439-433&0.1012&$127\pm35$&...&...&...&$136\pm30$&$124\pm15$&$713\pm23$&$490\pm51$&...\\
TXS 0454-220&0.4744&...&...&...&...&...&...&...&...&$47\pm8$\\
PHL 1598 & 0.4297&...&...&...&...&...&...&...&...&...\\
\tableline
\end{tabular}
\end{center}
\end{table}
\end{landscape}


\begin{landscape}
\begin{table}
\begin{center}
\caption{Results of Voigt profile fitting for lower ions in the $z=0.1602$ sub-DLA toward PHL 1226}
\vskip 1pt
\begin{tabular}{cccccccc}
\tableline\tableline
$ v^{1}$ 	& $b^{1}_{eff}$	&	$N_{\rm S II}$ 	&	$N_{\rm Si II}$	&	$N_{\rm Si II^{*}}$ 	&	$N_{\rm Fe II}$	&	$N_{\rm P II}$	&	$N_{\rm N II}$	\\
\tableline

-227.6		&	17.0	&	... 		&	$13.35\pm0.05$	&	...			&	$14.13\pm0.37$	&	$12.95\pm0.26$	&	$13.92\pm0.12$	\\
-180.2		&	24.7	&	$14.39\pm0.18$	&	$14.52\pm0.10$	&	...			&	...		&	...		&	$14.34\pm0.12$	\\
-145.7		&	19.1	&	$14.29\pm0.22$	&	$14.57\pm0.09$	&	...			&	$14.00\pm0.52$	&	...		&	$14.46\pm0.15$	\\
-111.7		&	20.6	&	$14.39\pm0.17$	&	$14.54\pm0.09$	&	...			&	$14.38\pm0.25$	&	$13.09\pm0.21$	&	$14.53\pm0.16$	\\
-85.4		&	7.0	&        ...		&	$12.37\pm0.33$	&	$12.42\pm0.16$		&	...		&	...		&	$13.45\pm0.34$	\\
-71.0		&	7.8	&        ...		&	$12.90\pm0.09$	&	...			&	$13.99\pm0.40$	&	...		&	$13.32\pm0.20$	\\
-50.7		&	11.0	&        ...		&	$13.43\pm0.13$	&	...			&	...		&	$12.76\pm0.33$	&	$13.99\pm0.14$	\\
-30.8		&	6.5	&        ...		&	$12.59\pm0.11$	&	...			&	...		&	...		&	$13.36\pm0.16$	\\
\tableline
\end{tabular}
\vskip 1pt
\begin{tabular}{ccccccc}
$ v^{1}$ 	& $b^{1}_{eff}$	& 	$N_{\rm C II}$ 		&	$N_{\rm S III}$	&	$N^{2}_{\rm Si III}$	&	$N_{\rm N I}$		&	$N_{\rm O I}$ 	\\
\tableline
-227.6		&	17.0	&		$>13.54$	&	$<13.65$	&	$>13.31$		&	$12.90\pm0.26$		&	$14.55\pm0.30$	\\
-180.2		&	24.7	&		$>14.81$	&	$<14.23$	&	$>13.64$		&	$12.97\pm0.39$		&	...		\\
-145.7		&	19.1	&		$>14.62$	&	$<14.29$	&	$>13.46$		&	$13.62\pm0.19$		&	...		\\
-111.7		&	20.6	&		$>14.66$	&	$<13.54$	&	$>13.55$		&	$14.05\pm0.09$		&	$>15.37$	\\
-85.4		&	7.0	&		$>14.18$	&	$<14.14$	&	$>13.08$		&	...			&	$14.54\pm0.39$	\\
-71.0		&	7.8	&		$>13.99$	&	$<14.56$	&	$>13.10$		&	$12.92\pm0.21$		&	...		\\
-50.7		&	11.0	&		$>14.09$	&	$<14.53$	&	$>13.37$		&	$12.70\pm0.36$		&	...		\\
-30.8		&	6.5	&		$13.85\pm0.41$	&	$<13.70$	&	$>12.40$		&	...			&	...		\\
-16.6		&	12.3	&		$13.81\pm0.23$	&	...		&	$12.44\pm0.09$		&	...			&	...		\\
\tableline
\end{tabular}

\tablenotetext{1}{$v$ and $b_{eff}$ denote the velocity and effective Doppler $b$ parameter in km s$^{-1}$. Column densities are logarithmic in cm$^{-2}$.}
\tablenotetext{2}{Additional absorption at -366.0, -301.9, -259.6 and 10.9 km s$^{-1}$ with $b_{eff}$ and log $N_{\rm Si III}$ values of 25.7, 12.6, 10.7, 18.0 km s$^{-1}$ and $12.92\pm0.04$,
$12.55\pm0.06$, $12.25\pm0.10$, $12.82\pm0.05$ cm$^{-2}$, respectively.}
\end{center}
\end{table}
\end{landscape}

\begin{landscape}
\begin{table}
\begin{center}
\caption{Results of Voigt profile fitting for higher ions in the $z=0.1602$ sub-DLA toward PHL 1226}
\vskip 5pt
\begin{tabular}{cccc}
\tableline\tableline
$ v^{1}$ & $b^{1}_{eff}$& 	$N_{\rm N V}$		&	$N_{\rm O VI}$	\\
\tableline
-367.4	&	36.9	&	$13.10\pm0.22$		&	$13.38\pm0.19$	\\
-293.5	&	6.4	&	...			&	$13.43\pm0.19$	\\
-237.5	&	20.1	&	$13.42\pm0.11$		&	$13.60\pm0.24$	\\
-154.8	&	62.9	&	$13.93\pm0.06$		&	$>14.96$	\\
-48.3	&	53.7	&	$13.51\pm0.12$		&	$>14.57$	\\
-48.1	&	44.5	&	$12.49\pm0.93$		&	$14.25\pm0.06$	\\
\tableline
\end{tabular}
\tablenotetext{1}{$v$ and $b_{eff}$ denote the velocity and effective Doppler $b$ parameter in km s$^{-1}$. Column densities are logarithmic in cm$^{-2}$.}
\end{center}
\end{table}
\end{landscape}

\begin{landscape}
\begin{table}
\begin{center}
\caption{Results of Voigt profile fitting for lower ions in the $z=0.1012$ sub-DLA toward PKS 0439-433}
\vskip 5pt
\begin{tabular}{cccccccc}
\tableline\tableline
$ v^{1}$ & $b^{1}_{eff}$&	$N_{\rm S II}$	&	$N_{\rm Si II}$	&	$N_{\rm Si II*}$&	$N_{\rm Fe II}$	&	$N_{\rm Fe III}$	&$N_{\rm S III}$	\\
\tableline
-125.1	&	28.1	&	...		&	$13.31\pm0.08$	&	$11.91\pm0.27$	&	$13.99\pm0.09$	&	$13.65\pm0.17$	&	$<13.82$		\\
-66.7	&	27.9	&	$14.68\pm0.05$	&	$>14.26$	&	$12.75\pm0.05$	&	$14.44\pm0.07$	&	$14.31\pm0.04$	&	$<14.50$		\\
-45.4	&	12.8	&	...		&	...		&	...		&	...		&	$14.04\pm0.08$	&	...			\\
-4.1	&	22.4	&	$14.77\pm0.04$	&	$>14.09$	&	$11.90\pm0.24$	&	$14.55\pm0.06$	&	$14.61\pm0.04$	&	$<15.09$		\\
50.2	&	16.4	&	...		&	$13.08\pm0.09$	&	...  		&	$13.77\pm0.07$	&	$14.11\pm0.05$	&	$<14.78$		\\
153.1	&	42.5	&	...		&	$13.67\pm0.07$	&	$12.48\pm0.09$	&	...		&	...		&	$<14.04$		\\
173.2	&	14.8	&	...		&	$12.93\pm0.31$	&	...		&	$13.72\pm0.07$	&	$13.53\pm0.12$	&	...			\\
218.4	&	18.0	&	...		&	$13.20\pm0.11$	&	...		&	...		&	...		&	...			\\
253.2	&	25.4	&	...		&	$13.10\pm0.10$	&	...		&	...		&	...		&	$<14.00$		\\
\tableline
\end{tabular}
\begin{tabular}{ccccccc}
\tableline\tableline
$v^{1}$ & $b^{1}_{eff}$	&	$N_{\rm P II}$	&	$N_{\rm N II}$	&	$N_{\rm Ar I}$	&	$N_{\rm N I}$	&	$N_{\rm Mn II}$	\\
\tableline
-125.1	&	28.1	&	...		&	$14.36\pm0.04$	&	... 		&	$13.94\pm0.06$	&	$13.33\pm0.19$	\\
-66.7	&	27.9	&	$12.97\pm0.11$	&	$15.25\pm0.16$	&	...		&	$14.55\pm0.06$	&	$12.96\pm0.39$	\\
-4.1	&      22.4	&	$13.13\pm0.08$	&	$14.70\pm0.06$	&	$13.27\pm0.15$	&	$14.51\pm0.07$	&	$13.15\pm0.25$	\\
50.2	&	16.4	&	...		&	$13.80\pm0.05$	&	...		&	...  		&	...		\\
153.1	&	42.5	&	...		&	$14.65\pm0.03$	&	...		&	$14.01\pm0.07$	&	...		\\
173.2	&	14.8	&	...		&	$13.64\pm0.11$	&	...		&	...		&	...		\\
218.4	&	18.0	&	...		&	$14.02\pm0.06$	&	...		&	$13.26\pm0.35$	&	$13.19\pm0.24$	\\
253.2	&	25.4	&	...		&	$13.36\pm0.20$	&	...		&	$13.00\pm0.51$	&	$13.48\pm0.16$	\\
\tableline
\end{tabular}
\vskip -15pt
\tablenotetext{1}{$v$ and $b_{eff}$ denote the velocity and effective Doppler $b$ parameter in km s$^{-1}$. Column densities are logarithmic in cm$^{-2}$.}
\end{center}
\end{table}
\end{landscape}

\begin{table}
\begin{center}
\caption{Results of Voigt profile fitting for higher ions in the $z=0.1012$ sub-DLA toward PKS 0439-433}
\vskip 5pt
\begin{tabular}{cccc}
\tableline\tableline
$v^{1}$ &	$b^{1}_{eff}$	&	$N_{\rm N V}$ 	&	$N_{\rm O VI}$	\\
\tableline
-115.3	&	38.7		&	$13.55\pm0.06$	&	$14.38\pm0.11$	\\
-46.6	&	33.4		&	$13.71\pm0.05$	&	$14.51\pm0.16$	\\
17.5	&	39.2		&	$13.67\pm0.05$	&	$14.38\pm0.12$	\\
122.0	&	44.7		&	$13.61\pm0.06$	&	$14.31\pm0.10$	\\
195.6	&	26.2		&	$13.56\pm0.06$	&	$14.25\pm0.15$	\\
243.2	&	22.2		&	$13.21\pm0.11$	&	$13.85\pm0.20$	\\
\tableline
\end{tabular}
\tablenotetext{1}{$v$ and $b_{eff}$ denote the velocity and effective Doppler $b$ parameter in km s$^{-1}$. Column densities are logarithmic in cm$^{-2}$.}
\end{center}
\end{table}

\begin{table}
\begin{center}
\caption{Results of Voigt profile fitting for ions in the $z=0.4744$ sub-DLA toward TXS 0454-220}
\begin{tabular}{ccccc}
\tableline\tableline
$v^{1}$ & $b^{1}_{eff}$	& 	$N_{\rm S II}$	&	$N_{\rm Si II}$	&	$N_{\rm C II}$	\\
\tableline
-75.0	&	14.2	&	$14.33\pm 0.09$	&	$13.07\pm0.06$	&	$>13.91$	\\
-41.9	&	8.5	&	$14.16\pm 0.14$	&	$>13.44$	&	$>14.44$	\\
-13.6	&	11.3	&	$14.63\pm 0.06$	&	$>13.83$	&	$>14.45$	\\
15.9	&	10.4	&	$14.44\pm0.07$	&	$>13.73$	&	$>14.32$	\\
42.9	&	8.1	&	...		&	$>13.51$	&	$>13.88$	\\
66.3	&	9.8	&	$14.00\pm0.16$	&	$>13.25$	&	$>14.18$	\\
99.6	&	9.5	&	...		&	$12.36\pm0.13$	&	$>13.24$	\\
\tableline
\end{tabular}
\begin{tabular}{ccccc}
\tableline\tableline
$v^{1}$ & $b^{1}_{eff}$	&	$N_{\rm N I}$			& 	$N_{\rm Ni II}$	&	$N_{\rm S III}$	\\
\tableline
-75.0	&	14.2 	&		...			&	... 		&	$<14.04$	\\
-41.9	&	8.5	&		...			&	$12.81\pm0.23$	&	$<13.87$	\\
-13.6	&	11.3	&	$13.83\pm0.09$			&	$13.21\pm0.10$	&	$<14.15$	\\
15.9	&	10.4	&	$12.95\pm0.39$			&	$12.93\pm0.19$	&	$<14.48$	\\
42.9	&	8.1	&		...			&	$12.65\pm0.35$	&	$<14.48$	\\
66.3	&	9.8	&	$12.99\pm0.17$			&	$12.80\pm0.30$	&	$<14.79$	\\
99.6	&	9.5	&		...			&	$12.81\pm0.23$	&	$<14.29$	\\
\tableline
\end{tabular}
\tablenotetext{1}{$v$ and $b_{eff}$ denote the velocity and effective Doppler $b$ parameter in km s$^{-1}$. Column densities are logarithmic in cm$^{-2}$.}
\end{center}
\end{table}

\begin{landscape}
\begin{table}
\begin{center}
\caption{Results of Voigt profile fitting for ions in the $z=0.4297$ sub-DLA toward PHL 1598 \label{phl1598velstruc}}
\begin{tabular}{cccccc}
\tableline\tableline
$v^{1}$	& $b^{1}_{eff}$	&	$N_{\rm S II}$	&	$N_{\rm Si II}$	&	$N_{\rm C II}$	&	$N_{\rm S III}$	\\
\tableline
-2.9	&	17.1	&	...		&	$12.92\pm0.05$	&	$13.87\pm0.04$	&	...		\\
23.1	&	7.0	&	...		&	$12.59\pm0.12$	&	$13.83\pm0.11$	&	...		\\
38.2	&	15.0	&	$14.31\pm0.05$	&	$12.97\pm0.05$	&	$13.89\pm0.07$	&	$<13.84$	\\
63.5	&	12.9	&	...		&	$12.39\pm0.21$	&	$13.35\pm0.10$	&	$<13.89$	\\
106.4	&	19.4	&	$13.37\pm0.33$	&	$12.41\pm0.12$	&	$13.48\pm0.07$	&	$<14.41$	\\
\tableline
\end{tabular}
\tablenotetext{1}{$v$ and $b_{eff}$ denote the velocity and effective Doppler $b$ parameter in km s$^{-1}$. Column densities are logarithmic in cm$^{-2}$.}
\end{center}
\end{table}
\end{landscape}


\begin{table}
\begin{center}
\caption{Total Column Densities for the $z=0.1602$ Absorber toward PHL 1226}
\vskip 10pt
\begin{tabular}{ccc}
\tableline\tableline
Ion&log $N^{fit}$&log $N^{AOD}$\\
&(cm$^{-2}$)&(cm$^{-2}$)\\
\tableline
H I &$19.48\pm0.10$&...\\
S II&$14.84 \pm 0.11$&$14.81 \pm 0.07$\\
S III&$<15.12$&...\\
Si II&$15.05 \pm 0.05$&$15.13 \pm 0.07$\\
Si II*&$12.42\pm0.16$&$12.41\pm0.14$\\
Si III&$>14.31$&$>14.27$\\
Fe II&$14.76\pm0.18$&$14.77\pm0.15$\\
Mn II&$<12.58$&...\\
C II&$> 15.31$&$> 15.28$\\
P II&$13.43\pm0.15$&$13.60\pm0.16$\\
N I&$14.27\pm0.08$&...\\
N II&$15.04 \pm 0.07$&$15.04 \pm 0.01$\\
N V&$14.20\pm0.05$&$14.25\pm0.13$\\
O I&$> 15.48$&$> 15.51$\\
O VI&$>15.19$&$>15.20$\\
\tableline
\end{tabular}
\end{center}
\end{table}

\begin{table}
\begin{center}
\caption{Total Column Densities for the $z=0.1012$ Absorber toward PKS 0439-433}
\vskip 10pt
\begin{tabular}{ccc}
\tableline\tableline
 Ion&log $N^{fit}$&log $N^{AOD}$\\
&(cm$^{-2}$)&(cm$^{-2}$)\\
\tableline
H I&$19.63\pm0.15$&...\\
S II&$15.03 \pm 0.03$&$15.02 \pm 0.06$\\
S III&$<15.38$&...\\
Si II&$>14.62$&$>14.54$\\
Si II*&$13.01\pm0.05$&$13.04\pm0.12$\\
Fe II&$14.92 \pm0.03$&$14.83\pm0.06$\\
Fe III&$14.97\pm0.02$&$15.05\pm0.04$\\
Mn II&$13.65\pm0.15$&$13.75\pm0.11$\\
P II&$13.36\pm0.06$&$13.43\pm0.14$\\
Ar I &$13.27\pm0.15$&$13.19\pm0.13$\\
N I&$14.95\pm0.04$&...\\
N II&$> 15.51$&$> 15.41$\\
N V&$14.36\pm0.02$&$14.34\pm0.06$\\
O VI &$15.10\pm0.06$&$15.14\pm0.03$\\
\tableline
\end{tabular}
\end{center}
\end{table}

\begin{table}
\begin{center}
\caption{Total Column Densities for the $z=0.4744$ Absorber toward TXS 0454-220}
\vskip 10pt
\begin{tabular}{ccc}
\tableline\tableline
Ion&log $N^{fit}$&log $N^{AOD}$\\
&(cm$^{-2}$)&(cm$^{-2}$)\\
\tableline
H I&$19.45 \pm 0.03$&...\\
S II&$15.06 \pm 0.04$&$14.98\pm 0.05$\\
S III&$<15.24$&...\\
Si II&$> 14.33$&$> 14.13$\\
Si II$^{*}$&$<11.86$&...\\
C II&$>15.04$&$> 14.92$\\
C II$^{*}$&$<12.57$&...\\
N I&$13.94\pm0.09$&$13.89\pm0.04$\\
Ni II&$13.69\pm0.08$&$13.63\pm0.07$\\
\tableline
\end{tabular}
\end{center}
\end{table}

\begin{table}
\begin{center}
\caption{Total Column Densities for the $z=0.4297$ Absorber toward PHL 1598}
\vskip 10pt
\begin{tabular}{ccc}
\tableline\tableline
Ion&log $N^{fit}$&log $N^{AOD}$\\
&(cm$^{-2}$)&(cm$^{-2}$)\\
\tableline
H I&$19.18 \pm 0.03$&...\\
S II&$14.36 \pm 0.05$&$14.34 \pm 0.10$\\
S III&$< 14.60$&...\\
Si II&$13.42 \pm 0.04$&$13.35 \pm 0.01$\\
C II &$14.43 \pm 0.04$&$14.36 \pm 0.02$\\
C II$^{*}$&$<12.85$&...\\
\tableline
\end{tabular}
\end{center}
\end{table}

\begin{table}
\begin{center}
\caption{Measured Element Abundances Relative to Solar}
\vskip 10pt
\begin{tabular}{ccccc}
\tableline\tableline
 Quasar&$z_{abs}$&log $N_{\rm{H I}}$&Element&[X/H]$^{1}$\\
 &&(cm$^{-2}$)&&\\
\tableline
PHL 1226&0.1602&$19.48\pm0.10$ &S&$0.24\pm0.15$\\
&&&Si&$0.06\pm0.11$\\
&&&Fe&$-0.22 \pm 0.21$\\
&&&P&$0.54\pm0.18$\\
&&&C&$> -0.60 $\\
&&&N&$-1.04\pm0.13$\\
&&&O&$> -0.69 $\\
PKS 0439-433&0.1012&$19.63 \pm 0.15$&S&$0.28\pm0.15$ \\
&&&Si&$>-0.52$\\
&&&Fe&$-0.21 \pm 0.15$\\
&&&Mn&$0.59\pm0.21$\\
&&&P&$0.32\pm0.16$\\
&&&Ar&$-0.76\pm0.21$\\
&&&N&$-0.51\pm0.16$ \\
TXS 0454-220&0.4744&$19.45 \pm 0.03$&S&$0.49\pm0.04$ \\
&&&Si&$ > -0.78$\\
&&&C&$>-0.88$\\
&&&N&$-1.34\pm0.09$\\
&&&Ni&$0.02\pm0.08$\\
PHL 1598&0.4297&$19.18 \pm 0.03$&S&$0.06\pm 0.06$\\
&&&Si&$-1.27 \pm 0.05$\\
&&&C&$-1.18 \pm 0.05$\\
\tableline
\end{tabular}
\tablenotetext{1}{Abundance estimates based on the dominant metal ionization state and H I. See text for effect of ionization corrections.}
\end{center}
\end{table}
\clearpage


\begin{table}
\begin{center}
\caption{Published Ionization Corrections for sub-DLAs \label{iontable}}
\vskip 10pt
\begin{tabular}{ccccccc}
\tableline
\tableline
Quasar		&	z$_{abs}$	&	log N$_{\rm H I}$	&f$^{a}_{[X/H]}$	&	$\epsilon$$^{b}$	&	Reference$^{c}$	&	 adopted $\epsilon$	\\
		&			&	(cm$^{-2}$)		&			&				&			&				\\
\tableline
Q1553+3548	&	0.083		&	19.55$\pm$0.15		&	0		&	[-0.30,  -0.12]		&	1		&	-0.21			\\
PKS 0439-433	&	0.1012		&	19.63$\pm$0.15		&	1		&	-0.18			&	2		&	-0.18			\\
Q0928+6025	&	0.1538		&	19.35$\pm$0.15		&	0		&	[-0.17, -0.11]		&	1		&	-0.14			\\
PHL 1226	&	0.1602		&	19.48$\pm$0.10		&	1		&	[-0.29, -0.24]		&	2		&	-0.26			\\
Q1435+3604	&	0.2026		&	19.80$\pm$0.10		&	1		&	[-0.11, 0.00]		&	1		&	-0.06			\\
Q0925+4004	&	0.2477		&	19.55$\pm$0.15		&	0		&	[-0.25, -0.16]		&	1		&	-0.21			\\
Q1001+5944	&	0.3035		&	19.32$\pm$0.10		&	0		&	[-0.28, -0.21]		&	1		&	-0.25			\\
PHL 1598	&	0.4297		&	19.18$\pm$0.03		&	1		&	-0.46$< $		&	2		&	-0.46			\\
TXS 0454-220	&	0.4744		&	19.45$\pm$0.03		&	1		&	-0.20$< $		&	2		&	-0.20			\\
Q1009-0026	&	0.8866		&	19.48$\pm$0.05		&	2		&	-0.15			&	4		&	0.15			\\
Q0826-2230	&	0.911		&	19.04$\pm$0.04		&	2		&	-0.15			&	4		&	0.15			\\
Q1311-0120	&	1.762		&	20.00$\pm$0.08		&	2		&	[0.10, 0.18]		&	3		&	0.14			\\
Q1103-2654	&	1.839		&	19.52$\pm$0.04		&	1		&	-0.31			&	3		&	-0.31			\\
Q2123-0050	&	2.058		&	19.35$\pm$0.10		&	2		&	[0.54, 0.63]		&	3		&	0.59			\\
Q1039-2719	&	2.139		&	19.55$\pm$0.15		&	2		&	[0.45, 0.51]		&	3		&	0.48			\\
Q1551+0908	&	2.32		&	19.70$\pm$0.05		&	1		&	-0.16			&	3		&	-0.16			\\
\tableline
\end{tabular}
\tablenotetext{a}{Type of metallicity measurement: 0 = S upper limit; 1 = S detection, 2 = Zn detection}
\tablenotetext{b}{Ionization correction $\epsilon (X) = [X/H]_{total} - [X_{dominant \, ion}/H I]$. A reported range in $\epsilon$ is denoted by numbers within square brackets.}
\tablenotetext{c}{References for ionization corrections: (1) \cite{Bat12}; (2) This study; (3) \cite{Som13}; (4) \cite{Mei07}.}
\end{center}
\end{table}
\clearpage

\begin{table}
\begin{center}
\caption{Velocity width measurements for the absorbers in our sample.\label{kinematicstable}}
\vskip 10pt
\begin{tabular}{ccccc}
\tableline
\tableline
Quasar		&	z$_{abs}$	&		[S/H]		&	$\Delta {v}_{90}$		&		Selected	\\
		&			&				&	(km$s^{-1}$)			&	Transition line		\\
\tableline						
PHL1226		&	0.1602		&	$+$0.24$\pm$0.15	&			152		&	Fe II $\lambda$ 1082	\\
PKS 0439-433	&	0.1012		&	$+$0.28$\pm$0.15	&			275		&	P II $\lambda$ 1153	\\
TXS 0454-220	&	0.4744		&	$+$0.49$\pm$0.04	&			155		&	S II $\lambda$ 1250	\\
PHL 1598	&	0.4297		&	$+$0.06$\pm$0.06	&			92		&	S II $\lambda$ 1259	\\
\tableline
\end{tabular}
\end{center}
\end{table}
\clearpage

\end{document}